\documentclass[11pt]{article}
\usepackage[dvips]{graphicx}
\usepackage{amssymb}
\usepackage{amsmath}
\usepackage{amssymb}
\usepackage{epsf}
\input{epsf}
\begin{document}
\begin{titlepage}
\title{\bf Variational Principle of Bogoliubov and Generalized Mean Fields in Many-Particle Interacting Systems\thanks{International Journal of Modern Physics B, 
\textbf{29}, 1530010 (63 pages) (2015); DOI: 10.1142/S0217979215300108}}  
\author{A. L. Kuzemsky 
\\
{\it Bogoliubov Laboratory of Theoretical Physics,} \\
{\it  Joint Institute for Nuclear Research,}\\
{\it 141980 Dubna, Moscow Region, Russia.}\\
{\it E-mail:  kuzemsky@theor.jinr.ru} \\
{\it http://theor.jinr.ru/\symbol{126}kuzemsky}}
\date{}
\maketitle
\begin{abstract}
The approach to the theory of many-particle interacting systems from a unified standpoint, based on the
variational principle for free energy  is reviewed. 
A systematic discussion is given of the approximate free energies of complex statistical systems.
The analysis is centered around  the
variational principle of N. N. Bogoliubov for free energy in the context of its applications to various 
problems of statistical mechanics and condensed matter physics.
The review presents a terse discussion  of   selected  works  carried out over the past few decades on the
theory of  many-particle interacting systems in terms of the  variational inequalities.
It is the purpose of this paper   to discuss some of the general principles which form the mathematical background to 
this approach, and to establish a connection of the variational technique with other methods,
such as the method of the mean (or self-consistent) field in the many-body problem, in which the effect of all 
the other particles on any given particle is approximated by a single 
averaged effect, thus reducing a many-body problem to a single-body problem.
The method is illustrated by applying it to various systems of many-particle interacting systems, such 
as Ising and Heisenberg models, superconducting and superfluid systems, strongly correlated systems, etc.
It seems likely that these technical advances in the many-body problem will 
be useful in suggesting new methods for treating and understanding many-particle interacting systems.
This work proposes a new, general and pedagogical presentation, intended both for those who are interested in  
basic aspects, and for those who are  interested in concrete applications.
\vspace{1cm}

\noindent \textbf{Keywords}: Statistical physics; mathematical physics; Helmholtz free energy; variational methods;
variational principle of N. N. Bogoliubov; Bogoliubov inequality; many-particle interacting systems;  cooperative phenomena; 
Bogoliubov theory of superfluidity; BCS-Bogoliubov microscopic theory of superconductivity; generalized mean fields; model Hamiltonians of
many-particle interacting systems; Ising, Heisenberg and Hubbard models.\\ 
\vspace{0.5cm}

\noindent\textbf{PACS}:  05.30.-d, 05.30.Fk, 05.30.Jp, 05.70.-a, 05.70.Fh, 02.90.+p\\\\
%
%
\end{abstract}
\end{titlepage}
\newpage
\tableofcontents
\newpage
%
%
%
%
%
%
%
%
%
\section{Introduction} 
%
%
%
%
The fundamental works of N.N. Bogoliubov on many-body theory and quantum field theory~\cite{bogocp12,nbog94,bb09,bb14}, on the theory of phase transitions,
and on the general theory of interacting systems provided a new perspective in various fields of mathematics and physics.
The    variational principle of N. N. Bogoliubov~\cite{bogocp12,nbog94,bb09,bb14,tyab} is a useful working tool and has been widely applied to many problems
of physical interest. 
It has a well-established place in the many-body theory  and
condensed matter physics~\cite{gira62,gira64,ishi68,okub71,huber68,huber73,mazo74,gira90,girar90}. 
The    variational principle of N. N. Bogoliubov has led to a better understanding
of various physical phenomena such as  superfluidity~\cite{bogocp12,nbog94,bb09,bb14}, 
superconductivity~\cite{bogocp12,nbog94,bb09,bb14,petqsm95}, phase transitions~\cite{bogocp12,nbog94,bb09,bb14,petqsm95,brent14}
and other cooperative phenomena~\cite{tyab,petqsm95,kuz09,kuz10}, etc.\\ 
Variational methods in physics and applied mathematics were formulated long 
ago~\cite{yurg68,blan92,gould95,basd07,michi09,berd09,maup74,gibbs,hill56,dru99}. 
It was Maupertuis~\cite{maup74}, who wrote in 1774 the celebrated statement:
\begin{quote}
\label{ } \em Nature, in the production of its effects, does so always by simplest means.
\end{quote}
Since that time variational methods have become an increasingly popular tool in mechanics, hydrodynamics, theory of
elasticity, etc. Moreover, the variational methods are   useful and workable tools for many areas of the quantum  
theory of atoms and molecules~\cite{gould95,pomr67,epst74,nesbet,ohmu60}, statistical many-particle physics and  condensed matter physics.
The variational methods have been applied widely in quantum-mechanical 
calculations~\cite{gould95,ohmu60,pomr67,epst74,nesbet,ohmu60},  in theory of  many-particle interacting 
systems~\cite{gira62,gira64,ishi68,okub71,huber68,huber73,mazo74,gira90,girar90} and theory of transport processes~\cite{kuz07,kuz11}.
As a result of these efforts  many important and effective methods were elaborated by various researchers.\\ 
From the other hand
the study of the quasiparticle excitations in many-particle systems
has been one of the most fascinating subjects for many years~\cite{petqsm95,tyab,kuz09,kuz10}.
The quantum field-theoretical techniques
have been widely applied to the statistical treatment of a large number of interacting
particles. Many-body calculations are often done for model systems of statistical
mechanics using the perturbation expansion. The basic procedure in many-body theory is
to find the relevant unperturbed Hamiltonian and then take into account the small perturbation
operator. This procedure, which works well for the weakly interacting
systems, needs the suitable reformulation for the many-body systems with complicated
spectra and strong interaction.\\
The considerable
progress in studying the spectra of elementary excitations and thermodynamic
properties of many-body systems has been for the most part due to the development
of the temperature-dependent Green  functions methods~\cite{petqsm95,tyab,kuz09,kuz10}.  The very important concept of the whole
method is the concept of  the generalized mean field~\cite{kuz09,kuz78,kuznc94,kuzrnc02}. These generalized mean fields have a complicated structure
for the strongly correlated case and are not reduced to the functional of the mean
densities of the electrons. The concept of the generalized mean fields and the relevant algebra of operators from which 
the corresponding Green  functions are constructed are the central ones to our treatment of the strongly interacting 
many-body systems.\\
It is the purpose of this paper  to discuss some of the general principles which form the physical and mathematical background to 
the variational approach, and to establish the connection of the variational technique with other methods in the theory of many-body problem. 
%
%
\section{The Variational Principles of Quantum Theory} 
%
%
It is well known that in quantum mechanics the eigenfunction $\psi_{i}$ of the lowest state of any system has the
property of making the integral
\begin{equation}\label{2.int}
 \int  \psi^{*}_{i}  H  \psi_{i}  d^{3}r
\end{equation}
a minimum. The value of integral is the corresponding eigenvalue $E_{i}$ of the Hamiltonian $H$  of a system. These circumstances
lead to a specific approximate method  (the variational method) of finding  $\psi_{i}$ and  $E_{i}$ by minimizing integral (\ref{2.int}) among a restricted
class of functions.\\
The variational method~\cite{gould95,ohmu60,pomr67,epst74,nesbet} enables one to make estimates of energy levels by using trial  wave functions $\psi_{T}$
\begin{equation}
   E_{T} = \frac{\int  \psi^{*}_{T}  H  \psi_{T}  d^{3}r}{ \int \psi^{*}_{T} \psi_{T}  d^{3}r}.
\end{equation}
The ground state $E_{0}$  gives the lowest possible energy the system can have. Hence, for the approximation of the ground state energy one
would like to minimize the expectation value of the energy with respect to a trial wave function.\\
In other words the variational principle states that the ground-state energy
of a quantum mechanical system is less than or equal to the expectation value
of the Hamiltonian with an arbitrary wave function. Given a trial wave function
with adjustable parameters, the best values of the parameters are those which
minimize the expectation value of the Hamiltonian.  The variational principle
consists in adjusting the available parameters, so as to maximize this lower bound.\\
An important method of finding approximate ground state energies and wave functions 
is called by the Rayleigh-Ritz variational principle~\cite{gould95,pomr67,epst74,nesbet}. 
The Rayleigh-Ritz variational principle for the ground state energy is the starting point of many computations 
and approximations in quantum mechanics and quantum chemistry of atoms and molecules. 
This principle states that the expectation value of $H$
in any state $|\psi \rangle$ is always greater than or equal to the ground state energy, $E_{0}$
\begin{equation}
   \frac{\langle \psi|H|\psi \rangle}{\langle \psi| \psi \rangle} \geq E_{0},
\end{equation}
or
\begin{equation}
\langle H  \rangle  \geq \langle \psi| H |\psi \rangle \geq E_{0}.
\end{equation}
Here $| \psi \rangle \in  \mathcal{G}$ is arbitrary pure quantum state and $H$ is   a Hamiltonian acting on a 
Hilbert space $\mathcal{G}$.
This relation becomes equality only when $\psi = \psi_{0}.$ Thus this principle gives the upper bound
to the ground state energy.\\
It will be instructive also to remind how the variational principle of quantum mechanics complements the
perturbation theory~\cite{hush94,deco04}. For this aim let us   consider  the  Rayleigh-Schr\"{o}dinger perturbation expansion.
The second-order level-shift $E^{0}_{2}$ of the ground state of a system  have the form
\begin{equation}
  E^{0}_{2} =  \sum_{j \neq 0} \frac{\langle \psi^{0}| V |\psi_{j}\rangle  \langle \psi_{j}| V |\psi^{0}\rangle}{(E^{0}  - E_{j})}
= \sum_{j \neq 0} \frac{| V_{0j}|^{2}}{(E^{0}  - E_{j})},
\end{equation}
where $V_{0j} = \langle \psi^{0}| V_{0j} |\psi_{j}\rangle$  and $|\psi^{0}\rangle$ is the unperturbed ground state.
It is clear then that $E^{0}_{2}$ is always negative.\\
The variational principle of quantum mechanics states that the ground state energy $E^{0}$ for the total Hamiltonian
$H$ is the minimum of the energy functional
\begin{equation}
   E \{\Psi \} = \langle \Psi| H |\Psi\rangle,
\end{equation}
where $\Psi$ is a trial wave function. It should be noted that it is possible to establish that the sum of all the
higher-order level shifts $E^{0}_{n},$ starting with $n = 2$, will be negative, providing that the relevant
perturbation series will converges to $E^{0}$.\\
To confirm this statement let us consider again the Hamiltonian
\begin{equation}
   H = H_{0} + \lambda V.
\end{equation}
It is reasonable to suppose that the ground state energy $E^{0} = E^{0}(\lambda)$ and the ground 
state $\Psi = \Psi (\lambda)$ of the Hamiltonian $H$ are analytic functions (at least for small $\lambda $).
Note that when one considers the many-body problem, the concept of relative boundedness is of use,
where a perturbation $\lambda V$ is small compared to $H_{0}$ in the sense that
$$(H_{0})^{2} \geq (\lambda^{2} V^{2}).$$
This means simply that the eigenvalues of the operator $((H_{0})^{2} - (\lambda^{2} V^{2}))$ are non-negative.\\
Then the corresponding perturbation  expansion  may be written in the form
\begin{equation}
   E_{0} = E^{(0)}_{0}  + \lambda E^{(1)}_{0} +   \lambda^{2} E^{(2)}_{0}  +   \lambda^{3} E^{(3)}_{0}    +  \quad  \ldots,
\end{equation}
where $E^{0}_{0} = \langle \psi^{0}| H |\psi^{0} \rangle$ and $E^{1}_{0} = \langle \psi^{0}| V |\psi^{0} \rangle$.\\
The variational approach states that
\begin{equation}
  E_{0} =  \textrm{min}  \left (\langle \Psi| H_{0} + \lambda V |\Psi \rangle \right ).
\end{equation}
Thus we obtain
\begin{eqnarray}
   \lambda^{2} E^{(2)}_{0}  +   
\lambda^{3} E^{(3)}_{0}    +  \quad  \ldots = E_{0} -  ( E^{(0)}_{0} + \lambda E^{(1)}_{0}) = \nonumber \\
\Bigl ( \textrm{min} \{\langle \Psi| H_{0} + \lambda V |\Psi \rangle \} - \langle \psi^{0} | H_{0} + \lambda V |\psi^{0}  \rangle \Bigr ).
\end{eqnarray}
In this expression the second part must satisfy the condition
\begin{equation}
  \Bigl ( \textrm{min} \{\langle \Psi| H_{0} + \lambda V |\Psi \rangle \} - \langle \psi^{0} | H_{0} + \lambda V |\psi^{0}  \rangle \Bigr ) \leq 0.
\end{equation}
In addition, in general case the relevant ground state $\Psi$  which yields a minimum  will not coincide with $\psi^{0}$.
Thus we obtain
\begin{equation}
    \lambda^{2} E^{(2)}_{0}  +   \lambda^{3} E^{(3)}_{0}    +  \quad  \ldots < 0.
\end{equation}
The last inequality can be rewritten as
\begin{equation}
   E^{(2)}_{0} < (\lambda E^{(3)}_{0} + \lambda^{2} E^{(4)}_{0} +  \quad  \ldots).
\end{equation}
In the limit $\lambda \rightarrow 0$ we have that $E^{(2)}_{0} < 0$. Thus the variational principle of quantum mechanics
confirms the results of the perturbation theory~\cite{fern01}.\\
It is worth mentioning that the Rayleigh-Ritz variational method has a long and interesting 
history~\cite{leis05,ilank09,leis09}.
Rayleigh's classical book \emph{Theory of Sound} was first published in 1877. In it are many examples of calculating 
fundamental natural frequencies of free vibration of continuum systems (strings, bars, beams, membranes, plates) by 
assuming the mode shape, and setting the maximum values of potential and kinetic energy in a cycle of motion equal to 
each other. This procedure is well known as \emph{Rayleigh's Method}.  In 1908, Ritz laid out his famous 
method for determining frequencies and mode shapes, choosing multiple admissible displacement functions, and minimizing 
a functional involving both potential and kinetic energies. He then demonstrated it in detail in 1909 for the completely 
free square plate. In 1911, Rayleigh wrote a paper congratulating Ritz on his work, but stating that he himself had 
used Ritz's method in many places in his book and in another publication.\\ 
Subsequently, hundreds of research articles and many books have appeared which use the method, some calling it the 
"Ritz method" and others the "Rayleigh-Ritz method." The   article~\cite{leis05} examined the method in detail, 
as Ritz presented it, and as Rayleigh claimed to have used it. A. W. Leissa~\cite{leis05}  concluded that, although Rayleigh did solve a few 
problems which involved minimization of a frequency, these solutions were not by the 
straightforward, direct method presented by Ritz and used subsequently by others. Therefore, Rayleigh's name 
should not be attached to the method. Additional informative comments were carried out in Refs.~\cite{ilank09,leis09}
%
%
%
%
%
\section{The Helmholtz Free Energy  and Statistical Thermodynamics} 
%
Variational methods in thermodynamics and statistical mechanics have been used widely since 
J. W. Gibbs  groundbreaking works~\cite{gibbs,hill56,dru99}. According to Gibbs  approach
a  workable procedure for the development of the statistical mechanical ensemble theory is to introduce the Gibbs
entropy postulate. 
Hence, as a result of the Gibbs ensemble method, the entropy $S$ can be 
expressed  in the form  of an average for all the ensembles, namely,
\begin{equation} \label{2.gen}
  S (N, V, E) =  - k_{B} \sum_{i}p_{i} \ln p_{i} = - k_{B} \Omega \Bigl ( \frac{1}{\Omega} \ln \frac{1}{\Omega} \Bigr ) = 
 k_{B}  \ln \Omega (N, V, E),
\end{equation}
where the summation over $i$ denotes a general   summation over all states of the system and $p_{i}$  is the probability of
observing state $i$ in the given ensemble and $k_{B}$ is the Boltzmann constant. This relation links entropy $S$ and
probability $p_{i}$.\\
It can be said that in this context the entropy is a state function which is according to the second law~\cite{hill56,hon3} is defined by
the relation
\begin{equation} 
  d S =  \beta (d E - d F).
\end{equation}
The energy $E$ and the Helmholtz free energy $F$ are the state functions~\cite{hill56,hon3}. The proportionality coefficient $\beta$
was termed as the \emph{thermodynamic temperature} ($\beta = 1/k_{B}T$) of the surrounding with which the system exchanges 
by heat $Q$ and work $W$.\\
Thus the postulate of 
equal probabilities in the microcanonical ensemble~\cite{kuz14} and  the Gibbs entropy postulate can be considered as a convenient 
starting points for the development of the statistical mechanical ensemble theory in a standard approach~\cite{hill56}. \\
After postulating  the entropy by means of Eq.(\ref{2.gen}), the thermodynamic equilibrium ensembles are determined by the
following criterion for equilibrium:
\begin{equation} 
 \left ( \delta S \right )_{E,V,N} =  0.
\end{equation}
This variational scheme is used for each ensemble (microcanonical, canonical and grand canonical) with different constraints
for each ensemble. In addition, this procedure introduces Lagrange multipliers which, in turn, must be identified with
thermodynamic intensive variables ($T, P$). From the other hand, the procedure of introducing
Lagrange multipliers and the task of identifying them with the thermodynamic intensive properties can be clarified
by invoking a more general criterion for thermodynamic equilibrium.\\
From the Gibbs entropy postulate, Eq.(\ref{2.gen}), the definitions of average and the normalization constraint
$\sum_{i}p_{i} = 1$ one obtains
\begin{eqnarray} 
  \delta S =  - k_{B} \sum_{i} (1 + \ln p_{i}) \delta p_{i},\\
  \delta E = \sum_{i} E_{i}  \delta p_{i},  \\
  \delta V = \sum_{i} V_{i}  \delta p_{i},  \\
 \sum_{i}   \delta p_{i} = 0.  
\end{eqnarray}
Using a Lagrange multiplier $\lambda$  together with the
variational condition, we obtain
\begin{equation} 
  \sum_{i}(E_{i} + P V_{i} + \lambda + k_{B} T + k_{B} T \ln p_{i} )\delta p_{i}  \geq 0.
\end{equation}
Here all $\delta p_{i}$ are considered as the independent variables. Thus we deduce that
\begin{equation} 
  p_{i} = \exp \Bigl ( - \beta \lambda - 1 - \beta (P V_{i} + E_{i})  \Bigr ), \quad \beta = (k_{B} T)^{-1}.
\end{equation}
The Lagrange multiplier $\lambda$, can be determined directly from the definition of entropy  (\ref{2.gen}).
\begin{eqnarray} 
S = - k_{B} \sum_{i} p_{i}  \Bigl ( \frac{ E_{i} + P V_{i} + \lambda + k_{B} T }{k_{B} T} \Bigr ) =
\frac{  \Bigl (E  + P V  + \lambda + k_{B} T  \Bigr )}{ T}.  
\end{eqnarray}
Thus we arrive at
\begin{eqnarray} 
  \lambda + k_{B} T  =  T S - E - P V = - G,\\
   p_{i} = \exp  \beta  \Bigl ( G - P V_{i} - E_{i}   \Bigr ).
   \label{2.prob}  
\end{eqnarray}
Here $G$ is the Gibbs energy (or Gibbs free energy). It may also be defined with the aid of the Helmholtz free energy
$G = H - T S$. Here $H (S, P, N)$ is the enthalpy~\cite{hon3}.\\ 
The usefulness of the thermodynamic potentials $G$ and $F$ may be clarified within the statistical thermodynamics~\cite{hill56}.
For the \emph{microcanonical} ensemble one should substitute $E_{i} = E$ and $V_{i} = V$, which are fixed for every
system and since $G - P V - E = S$, Eq.(\ref{2.prob}) becomes
\begin{equation} 
  p_{i} = e^{- S/k_{B}}.
\end{equation}
For the canonical ensemble one should substitute  $V_{i} = V$, which is given for each
system and in this case Eq.(\ref{2.prob}) can be written as
\begin{equation} 
  p_{i} = e^{\beta (F - E_{i})}.
\end{equation}
Here  $F = G - P V$  denotes  the \emph{Helmholtz free energy}. Thus the free energy $F$   is defined by
\begin{equation} 
  F = E - T S.
\end{equation}
The Helmholtz free energy describes an energy which is available in the form of useful work.\\
The second law of thermodynamics  asserts that in every neighborhood of any state $\mathcal{A}$ in an 
adiabatically isolated system there exist other states that are inaccessible from $\mathcal{A}.$ This statement 
in terms of the \emph{entropy} $S$ and \emph{heat} $Q$ can be
formulated as 
\begin{equation}
d S = d Q/T + d \sigma. 
\end{equation}
Thus the only states available in an adiabatic process  
($dQ = 0$, or $d S = d \sigma$) are those which lead to an increase of the entropy $S$. Here $d \sigma \geq 0$ defines the 
entropy production $\sigma$  due to the irreversibility of the transformation.\\
It is of use to analyze the expression
\begin{equation} 
 d F = d E - T d S - S d T = - S d T - T d \sigma - P d V + \sum \mu_{i} N_{i}.
\end{equation}
Free energy change $\Delta F$ of a system during a  transformation of a system describes a balance of the work exchanged
with the surroundings. If $\Delta F  > 0$,  $\Delta F$ represents the minimum work that must be incurred for the system 
to carry out the transformation. In the case $\Delta F < 0$, $|\Delta F|$ represents the maximum work that can be obtained 
from the system during the transformation. It is obvious that
\begin{equation} 
  d F = d E - T d \sigma -S d T.
\end{equation}
In a closed system without chemical reaction and in the absence of any other energy exchange, the variation 
$\Delta F = -S dT - T dS - P dV + \sum \mu_{i}N_{i}$ can be rewritten in the form
\begin{equation} 
  d F =  - T d \sigma  \leq 0.
\end{equation}
%
%
%
%
%
It means that function $F$ decreases and tends towards a minimum corresponding to equilibrium. Thus the
Helmholtz free energy is the  thermodynamic potential of a system subjected to the constraints constant $T, V, N_{i}$.\\
The \emph{Gibbs free energy} (free enthalpy) is defined by
\begin{equation} 
  G = H - T S = F + P V.
\end{equation}
The physical meaning of the Gibbs free energy is clarified when considering evolution of a system from a certain initial state
to a final state. The Gibbs free energy change $\Delta G$ then represents the work exchanged by the system with 
its environment and the work of the pressure forces, during a reversible transformation of the system.
Here $H = E + V P = T S + V P + \sum \mu_{i} N_{i}$ is the thermodynamic potential of a system termed by 
\emph{enthalpy}~\cite{hon3}.  The Gibbs free energy is the  thermodynamic potential of a system subjected to 
the constraints constant $T,P,N_{i}$. In this case
\begin{equation} 
  d G =  - T d \sigma \leq 0.
\end{equation}
Thus  in the closed system without chemical reaction and in the absence of any other energy exchange at constant temperature, 
pressure and amount of substance the function $G$ can only decrease and reach a minimum at equilibrium.\\
It will be of use to mention another class of thermodynamic potentials, termed by the Massieu-Planck functions. These
objects may be deduced from the fundamental relations in the entropy representations, $S = S (E, V, N)$. The
corresponding differential form may be written as
\begin{equation}
  d S = \frac{1}{T} d E +  \frac{P}{T} d V  - \frac{\mu}{T} d N.
\end{equation}
Thus the suitable variables for a Legendre transform will be $1/T$, $P/T$ and $\mu /T$. In
some cases working with these variables is more convenient.\\
It is worth noting that in terms of the Gibbs ensemble method
the free energy is the  thermodynamic potential of a system subjected to the constraints constant $T,V,N_{i}$.
Moreover,  the  thermodynamic potentials should be defined properly in the thermodynamic limit.
The problem of the thermodynamic limit in statistical physics was discussed in detail by A. L. Kuzemsky~\cite{kuz14}.
To clarify this notion, let us consider the logarithm of the partition function $Q  (\theta, V, N)$
\begin{equation}\label{3.2}
 F (\theta, V, N) =   - \theta  \ln  Q  (\theta, V, N).
\end{equation}
This expression determines the free energy $F$ of the system on the basis of canonical distribution. The standard way of
reasoning in the equilibrium statistical mechanics do not requires the knowledge of the exact value of the
function $F (\theta, V, N)$. For real system it is sufficient to know the thermodynamic (infinite volume) 
limit~\cite{petqsm95,hill56,kuz14,mef64}
\begin{equation}\label{3.3}
 \lim_{N \rightarrow \infty}   \frac{F (\theta, V, N)}{N}\vert_{V/N = \textrm{const}} =  f(\theta, V/N).
\end{equation}
Here $f(\theta, V/N)$ is the free energy per particle. It is clear that this function determines all the thermodynamic 
properties of the system.\\
Let us summarize  
the criteria for equilibrium briefly. In a system of constant $V$ and $S$,
the internal energy has its \emph{minimum} value, whereas in a system of constant $E$ and $V$, the entropy 
has its \emph{maximum} value.\\
It should be noted that the pair of independent variables $(V,S)$ is not suitable one because of the entropy 
is not convenient to measure or control. Hence it would be of use
to have fundamental equations with independent variables that is easier to control.
The two convenient choices are possible. First we take the $P$ and $T$ pair. From the practical point of view this is
a convenient pair of variables which are easy to control (measure). For systems with constant pressure the
best suited state function is the Gibbs free energy (also called free enthalpy)
\begin{equation}
 G = H - T S.
\end{equation}
Second relevant pair is $V$ and $T$.   For
systems with constant volume (and variable pressure), the suitable suited state function is
the Helmholtz free energy
\begin{equation}
 F = E - T S.
\end{equation}
Any state function can be used
to describe any system (at equilibrium, of course), but for a given system some are more
convenient than others.
The change of the Helmholtz free energy can be written as 
\begin{equation}
 d F = d E - T d S - S d T.
\end{equation}
Combining this equation with $dU = TdS - PdV$ we obtain the relation of the form
\begin{equation}
 d F =   - P d V - S d T.
\end{equation}
In terms of variables $(T, V)$ we find
\begin{equation}
 d F =  \Bigl (  \frac{\partial F}{\partial T} \Bigr )\Bigr |_{V} d T  +  \Bigl ( \frac{\partial F}{\partial V}  \Bigr ) \Bigr |_{T} d V.  
\end{equation}
Comparing the equations one can see that 
\begin{equation}
 S = - \Bigl (  \frac{\partial F}{\partial T} \Bigr )\Bigr |_{V}, \quad P = \Bigl ( \frac{\partial F}{\partial V}  \Bigr )\Bigr |_{T}.
\end{equation}
At constant $T$ and $V$ the equilibrium states corresponds to the
minimum of Helmholtz free energy $(d F = 0)$. From  $F = E - T S$
we may suppose that low values of $F$ are obtained with low values of $E$
and high values of $S$.\\
In terms of a general statistical-mechanical formalism~\cite{bb09,bb14,petqsm95}, a many-particle system with Hamiltonian
$H$ in contact with a heat bath at temperature $T$ in a state described by the statistical operator $\rho$ has a free energy
\begin{equation}
 F = \textrm{Tr} (\rho H) + k_{B} T \textrm{Tr} (\rho \ln \rho).
\end{equation}
The free energy takes its minimum value
\begin{equation}
 F_{\textrm{eq}} = - k_{B} T  \ln Z
\end{equation}
in the \emph{equilibrium} state characterized by the canonical distribution
\begin{equation}
 \rho_{\textrm{eq}} =   Z^{-1}  \exp (- H \beta); \quad Z =  \textrm{Tr} \exp (- H \beta). 
\end{equation}
Before turning to the next topic, an important  remark about the free energy  will not be out of place here. I. Novak~\cite{novak05}
attempted  to give a microscopic description of Le Chatelier's principle~\cite{hbcal} in statistical systems.
Novak have carried out    interesting analysis based on microscopic descriptors (energy levels and their populations) 
that provides visualization of free energies and conceptual rationalization of 
Le Chatelier's principle. The misconception "\emph{nature favors equilibrium}" was highlighted. This problem is a delicate 
one and requires a careful discussion~\cite{sear09}. Dasmeh et al. showed~\cite{sear09} that
Le Chatelier's principle states that when a system is disturbed, it will shift its equilibrium to counteract the disturbance. 
However for a chemical reaction in a small, confined system, the probability of observing it proceed in the opposite 
direction to that predicted by Le Chatelier's principle, can be significant. Their study provided a molecular level proof 
of Le Chatelier's principle for the case of a temperature change. Moreover, a new, exact mathematical expression was derived 
that is valid for arbitrary system sizes and gives the relative probability that a single experiment 
will proceed in the endothermic or exothermic direction, in terms of a microscopic phase function. They showed that 
the average of the time integral of this function is the maximum possible value of the purely irreversible entropy 
production for the thermal relaxation process. The results obtained were tested 
against computer simulations of the unfolding of a polypeptide. It was proven  that any equilibrium reaction 
mixture on average responds to a temperature increase by shifting its point of equilibrium in the endothermic direction.
%
%
\section{Approximate Calculations of Helmholtz Free Energy} 
%
%
Statistical mechanics provides effective and workable tools for describing the behavior of the systems of many 
interacting particles. One of such approaches for describing systems in equilibrium consists in evaluation the partition
function $Z$ and then the free energy.\\
Now we must take note of the different methods for obtaining the approximate Helmholtz free energy
in the theory of many-particle systems.  Roughly speaking, there are two approaches, namely the perturbation method  and
the variational method.\\
Thermodynamic perturbation theory~\cite{peier33,peier79,peierws,fern03} may be  applied to systems that undergo a phase transition. 
It was shown~\cite{herm68} that certain conditions are necessary in order 
that the application of the perturbation does not change the qualitative features of the phase transition. 
Usually, the shift in the critical temperature are determined to two orders in the perturbation parameter.
Let us consider here the perturbation method~\cite{herm68} very briefly.\\
In paper~\cite{herm68} authors considered a system with Hamiltonian $H_{0}$, that
undergoes a phase transition at critical temperature
$T_{C}^{0}$. The task was to determine for what class of perturbing potentials
$V$ will the system with Hamiltonian $H_{0} + V$
have a phase transition with qualitatively the same
features as the unperturbed system. In their paper
authors~\cite{herm68} studied that question using  thermodynamic
perturbation theory~\cite{peier33,peier79,peierws}.   They found that an
expansion for the perturbed thermodynamic functions
can be term by term divergent at the critical
temperature $T_{C}^{0}$ for a class of potentials $V$. Under
certain conditions the series can be resummed,
in which case the phase transition remains qualitatively
the same as in the unperturbed system
but the location of the critical temperature is shifted. \\
The starting point was the partition function $Z_{0}$ for a system whose Hamiltonian is
$H_{0}$
\begin{equation}
 Z_{0} = \textrm{Tr} \exp ( - H_{0} \beta). 
\end{equation}
For a system with Hamiltonian $H_{0} + \lambda V$, the partition
function $Z$ is given by
\begin{equation}
 Z  = \textrm{Tr}  \exp [ - (H_{0} + \lambda V)\beta].    
\end{equation}
It is possible to obtain formally  an expansion for $Z$ in
terms of the properties of the unperturbed system by expanding that part of the exponential containing
the perturbation in the following way~\cite{herm68} when $V$  and $H_{0}$ commute:
\begin{eqnarray}
  Z  = \textrm{Tr} \Bigl ( \exp ( - H_{0} \beta) \sum_{n}^{\infty} \frac{1}{n !} [- \lambda \beta]^{n} V^{n} \Bigr ) = 
 Z_{0} \sum_{n}^{\infty} \frac{1}{n !} [- \lambda \beta]^{n} \langle V^{n} \rangle_{0},
\end{eqnarray}
where
\begin{equation}
 Z_{0} \langle V^{n} \rangle_{0} = \textrm{Tr} \Bigl ( \exp [- H_{0} \beta]  V^{n} \Bigr ).
\end{equation}
Then the expression for $Z$ can be written as
\begin{equation}
 \frac{Z}{Z_{0}} = \exp  \Bigl ( \ln [ 1 +  \sum_{n}^{\infty} \frac{1}{n !} (- \lambda \beta)^{n}  \langle V^{n} \rangle_{0}  ]  \Bigr )
\end{equation}
The free energy per particle $f$ is given by
\begin{equation}
\beta  f_{p} = \beta  f_{0} - \frac{1}{N} \ln \Bigl ( 1 +  \sum_{n}^{\infty} \frac{1}{n !} (- \lambda \beta)^{n}  \langle V^{n} \rangle_{0}    \Bigr ),
\end{equation}
where $ f_{p}$ and  $f_{0}$ are the perturbed and unperturbed free energy per particle, respectively, and $N$ is the number
of particles in the system. The standard way to proceed consists of in expanding the logarithm in powers of $\lambda.$
As a result one obtains~\cite{herm68}
\begin{eqnarray}
 \beta  f_{p} = \beta  f_{0} + \frac{\lambda \beta}{N}  \langle V \rangle_{0} - \frac{\lambda^{2} \beta^{2}}{2 !} \frac{1}{N}  \Bigl ( \langle V^{2} \rangle_{0} - \langle V \rangle_{0}^{2} \Bigr )
\nonumber \\ + \frac{\lambda^{3} \beta^{3}}{3 !} \frac{1}{N}  \Bigl ( \langle V^{3} \rangle_{0} - 3 \langle V^{2} \rangle_{0} \langle V \rangle_{0}   + \langle V \rangle_{0}^{3} \Bigr ) + \ldots
\end{eqnarray}
To proceed, it is supposed usually that the thermodynamics of the unperturbed system is known and the perturbation series
(if they converge) may provide us with suitable corrections. If the terms in the expansion diverge, they may, in principle,
be regularized under some conditions. For example, perturbation expansions for the equation of state of a fluid whose
intermolecular potential can be regarded as consisting of the sum of a strong and weak part give reasonable qualitative
results~\cite{stor69,donald00}.\\
In  paper by Fernandes~\cite{fern03} author investigated the application of perturbation
theory to the canonical partition function of statistical mechanics.
The  Schwinger and Rayleigh-Schr\"{o}dinger perturbation theory were outlined and plausible arguments were formulated 
that both should give the same result. 
It was shown that by introducing adjustable parameters in the unperturbed
or reference Hamiltonian operator, one can improve the rate
of convergence of Schwinger perturbation theory.  The
same parameters are also suitable for Rayleigh-Schr\"{o}dinger
perturbation theory.  Author discussed also a possibility of   variational
improvements of perturbation theory and gave a simpler
proof of a previously derived result about the choice of the
energy shift parameter.  It was also shown that some variational
parameters correct the anomalous behavior of the partition
function at high temperatures in both Schwinger and
Rayleigh-Schr\"{o}dinger perturbation theories. 
It should be stressed, however, that 
the perturbation method is  valid for small perturbations only. The variational method is more
flexible tool~\cite{hush94,deco04,nmarc58,garr64,chang70,keste88,chen94,gray04} and in many cases is more appropriate in spite of the obvious shortcomings.  But both the methods are
interrelated deeply~\cite{deco04} and enrich each other.\\
R. Peierls~\cite{peier79,peierws,peier38}  pointed at the circumstance that for a many-particle system  in thermal equilibrium
there is a minimum property of the free energy which may be considered as a generalization of the variational principle
for the lowest eigenvalue in quantum mechanics.     
Peierls attracted attention to the fact that the free energy has a specific property which can be  formulated
in the following way. Let us consider an arbitrary set  of orthogonal and normalized  
functions $\{\varphi_{1}, \varphi_{2}, \ldots \varphi_{n}, \ldots \}.$ The expectation value of the Hamiltonian $H$ for
$n$th of them will be written as
\begin{equation}
H_{nn} = \int \varphi_{n}^{*} H \varphi_{n} d r.
\end{equation}
The statement is that for any temperature $T$ the function
\begin{equation}
 \tilde{F} = - k_{B}T \log \tilde{Z} = - k_{B}T \log \sum_{n}  \exp[ - H_{nn} \beta]
\end{equation}
which would represent the free energy if the $H_{nn}$ were the true eigenvalues, is \emph{higher} than the true
free energy
\begin{equation}
 F_{0} = - k_{B}T \log Z_{0} = - k_{B} T \log \sum_{n}  \exp[ - E_{n} \beta],
\end{equation}
or
\begin{equation}
  \tilde{F} \geq  F_{0}.
\end{equation}
This is equivalent to saying that the partition function, as formed by means of the expectation values $H_{nn}$:
\begin{equation}
 \tilde{Z} =  \sum_{n}  \exp[ - H_{nn} \beta]
\end{equation}
is less than the true partition function
\begin{equation}
 Z_{0} =  \sum_{n}  \exp[ - E_{n} \beta], 
\end{equation}
or
\begin{equation}
 Z_{0} =  \sum_{n}  \exp[ - E_{n} \beta] \geq   \tilde{Z} =  \sum_{n}  \exp[ - H_{nn} \beta]. 
\end{equation}
Peierls~\cite{peier38} formulated the more general statement, namely, that if  $f(E)$ is a function with the properties
\begin{equation}
 \frac{d f}{d E} < 0, \quad \frac{d^{2} f}{d E^{2}} > 0,
\end{equation}
the expression 
\begin{equation}
 f = \sum_{n} f(H_{nn})
\end{equation}
is less than
\begin{equation}
 f_{0} = \sum_{n} f(E_{n}).
\end{equation}
To summarize, Peierls has proved a kind of theorem a special case of which gives a lower bound to the
partition sum and hence an upper bound to the free energy of a quantum mechanical system 
\begin{equation}
 \sum_{k}  \exp[ - E_{k} \beta] \geq   \sum_{n}  \exp[ - H_{nn} \beta].
\end{equation}
When $\beta \to \infty$ the theorem is obvious, reducing to the fundamental inequality $E_{k} \leq H_{nn}$ for all $n$.
However for finite $\beta$ it is not so obvious since higher eigenvalues of $H$ do not necessarily lie lower than
corresponding diagonal matrix elements $H_{nn}$. T. D. Schultz~\cite{tdschu58}  skilfully remarked that, in fact, the
Peierls inequality does not depend on the fact that $\exp[ - E  \beta]$ is a monotonically decreasing function of $E$,
as might be  concluded from the original proof. It depends only on the fact that the exponential function is concave upward.
Schultz~\cite{tdschu58} proposed a simple proof of the theorem under this somewhat general condition.\\
Let ${\varphi_{n}}$ be a complete orthonormal set of state vectors and let $\mathbf{A}$ be an Hermitean operator which
for convenience is assumed to have a pure point spectrum with eigenvalues $a_{k}$ and eigenstates $\psi_{k}$. Let $f(x)$
be a real valued function such that
\begin{equation}\label{pineq2}
 \frac{d^{2} f}{d x^{2}} > 0
\end{equation}
in an interval including the whole spectrum of $a_{k}$. Then if $\textrm{Tr} f(\mathbf{A})$ exists it can be proven
that the following statement holds
\begin{equation}\label{pineq3}
 \textrm{Tr} f(\mathbf{A}) \geq \sum_{n} f(a_{nn}),
\end{equation}
where $a_{nn} = \langle n| \mathbf{A}|n \rangle$. The equality holds if and only if the ${\varphi_{n}}$ are the eigenstates
of $\mathbf{A}$. \\
Since
\begin{equation}
 \textrm{Tr} f(\mathbf{A}) = \sum_{n} \langle n| f(\mathbf{A})| n\rangle
\end{equation}
it is sufficient for the proof to point out that the relation (\ref{pineq3}) follows from
\begin{equation} \label{pineq4}
 \langle n| f(\mathbf{A})|n \rangle \geq f(a_{nn}),
\end{equation}
which is valid for all $n$. The inequalities (\ref{pineq4}) were derived from
\begin{equation}\label{pineq5}
 f(a_{k}) \geq f(a_{nn}) + (a_{k} - a_{nn})f'(a_{nn}),
\end{equation}
which is a consequence of equation (\ref{pineq2}), the right hand side for fixed $n$ being the line tangent to $f(a_{k})$
at $a_{nn}$. Multiplying (\ref{pineq5}) by $ | \langle n|k \rangle |^{2} $ and summing on $k$ one obtains (\ref{pineq4}).
Schultz~\cite{tdschu58} observed further that the equality in (\ref{pineq4}) holds if and only 
if  $ | \langle n|k \rangle |^{2} = 0$  unless $a_{k} = a_{nn}$, i.e. if and only if ${\varphi_{n}}$ is an eigenstate
of $\mathbf{A}$.\\
If $f(\mathbf{A})$ is positive definite, then the set ${\varphi_{n}}$ need not be complete, since the theorem is true even more strongly 
if positive terms are omitted from the sum $\sum_{n} f(a_{nn})$.\\
With the choice $f(\mathbf{A}) = \exp ( - \mathbf{A})$ and $\mathbf{A} = \mathbf{H} \beta$, the original theorem of Peierls
giving an upper bound to the free energy is reproduced. With $\mathbf{A} = (\mathbf{H} - \mu \mathbf{N}) \beta$ we have an 
analogous theorem for the grand potential. The theorem proved by Schultz~\cite{tdschu58} is a generalization in that it no longer requires
$f(x)$ to be monotonic; it requires only that $ \textrm{Tr} f(\mathbf{A})$ be finite which can occur even if $f(x)$
is not monotonic provided $\mathbf{A}$ is bounded.\\
Peierls variational theorem was discussed and applied in  a number of papers (see, e.g. Refs.~\cite{tdschu58,falk63,falk64,falk70}).
It has much more generality than, say, the A. Lidiard~\cite{lidi53}  consideration on a minimum property of the free energy.
Lidiard~\cite{lidi53} derived the approximate free energy expression in a way which
shows a strong analogy with the approximate Hartree method of quantum mechanics.
By his derivation he refined the earlier calculations made by Koppe an  Wohlfarth in the context of
description of the influence of the exchange energy on the thermal properties of free electrons in metals.  
%
%
%
%
%
\section{The Mean Field Concept}
%
In general case, a many-particle  system with interactions is  very difficult to solve exactly, except for special simple 
cases. Theory of molecular (or mean) field permits one to obtain an approximate solution to the problem.
In condensed matter physics, mean field theory (or self-consistent field theory) studies the behavior of large many-particle systems 
by studying a simpler models. The effect of all the other particles on any given particle is approximated by a single 
averaged effect, thus reducing a many-body problem to a single-body problem.\\
It is well known that molecular fields in various variants appear in the
simplified analysis of many different kinds of many-particle interacting systems.
The mean field concept was originally
formulated for many-particle systems (in an implicit
form) in Van der Waals~\cite{vderw3,hill48} dissertation "On the Continuity of Gaseous and Liquid States".  
Van der Waals conjectured that the volume correction to the equation of state
would lead only to a  trivial reduction of the available
space for the molecular motion by an amount $b$ equal to
the overall volume of the molecules. In reality, the
measurements  lead him to be much more complicated dependence.
He found that both the  corrections  should be taken into account.
Those were the volume correction $b$, and the pressure correction
$a/V^2$, which led him to the Van der Waals equation~\cite{hill48}.
Thus, Van der Waals came to conclusion that "the range of
attractive forces contains many neighboring molecules".
The equation derived by Van der Waals was similar to the ideal gas equation except that the pressure is
increased and the volume decreased from the ideal gas values.
Hence, the many-particle behavior was reduced to
\emph{effective} (or renormalized) behavior of a single particle in a medium (or a field). 
The later development of this line of reasoning led to the
fruitful concept, that it may be reasonable to describe approximately the complex many-particle
behavior of gases, liquids, and solids in terms
of a single particle moving in an average (or effective)
field created by all the other particles, considered as some homogeneous (or inhomogeneous) environment.\\
Later, these ideas were extended to the physics of magnetic phenomena~\cite{kuz09,tyab,smart,rscom08}, where
magnetic substances were considered as some kind of a specific liquid. 
This approach was elaborated in the physics of magnetism by P. Curie  and P. Weiss.
The mean field (molecular field) replaces the interaction of all the other particles to an arbitrary 
particle~\cite{palf02}. In the mean field approximation, the energy of a system is replaced by the sum of 
identical single particle energies that describe the interactions of each particle with an effective mean field.\\
Beginning from 1907 the Weiss molecular-field approximation 
became widespread in the theory of magnetic phenomena~\cite{kuz09,tyab,smart,rscom08},
and even at the present time it is still being used efficiently. Nevertheless, back in 1965 it
was noticed that~\cite{hbcal65}:
\begin{quote}
"\emph{The Weiss molecular field theory plays an enigmatic
role in the statistical mechanics of magnetism}".
\end{quote}
In order to explain the concept of the molecular
field on the example of the Heisenberg ferromagnet
one has to transform the original many-particle
Hamiltonian 
\begin{equation}
\label{13.hh} H = -  \sum_{ij}  J(i-j) \mathbf{S_{i}}  \mathbf{S_{j}} 
-g\mu_{B}H\sum_{i}S_{i}^{z}.
\end{equation}
into the following reduced one-particle
Hamiltonian
\begin{equation} 
\mathcal{H}  = -   2 \mu_{0} \mu_{B} \vec{\textbf{S}} \cdot \vec{\textbf{h}}^{(mf)}.  \nonumber
\end{equation}
The coupling coefficient $J(i-j)$ is the measure of the  exchange
interaction between  spins at the lattice sites $i$ and $j$ and
is defined usually to have the property J(i - j = 0) = 0.
This transformation was achieved with the help of the
identity~\cite{kuz09,tyab,smart,rscom08}
$$\vec{S} \cdot \vec{S'} = \vec{S}\cdot \langle \vec{S'}\rangle + \langle \vec{S} \rangle \cdot \vec{S'} - 
\langle \vec{S} \rangle \cdot \langle \vec{S'}\rangle + C.$$
Here, the constant $C = (\vec{S} - \langle \vec{S} \rangle) \cdot (\vec{S'} - \langle \vec{S'} \rangle)$ describes
spin correlations. The usual molecular-field approximation
is equivalent to discarding the third term in the
right hand side of the above equation, and using the
approximation $C \sim \langle C \rangle = 
\langle \vec{S} \cdot \vec{S'} \rangle - \langle \vec{S} \rangle \cdot \langle \vec{S'}\rangle.$ for the
constant $C$. \\ 
There is large diversity of the mean-field theories adapted to various concrete applications~\cite{kuz09,tyab,smart,rscom08}.\\
Mean field theory has been applied to a number of models of physical systems so as to study phenomena such as phase 
transitions~\cite{honi99,sole11}. One of the first application was Ising model~\cite{kuz09,tyab,smart,rscom08}.
Consider the Ising model on an $N$-dimensional cubic lattice. The Hamiltonian is given by
\begin{equation}
H = -J \sum_{\langle i,j\rangle} S_i S_{j} - h \sum_i S_i,  
\end{equation}
where the $\sum_{\langle i,j\rangle}$ indicates summation over the pair of nearest neighbors $\langle i,j\rangle$ , 
and $S_i = \pm 1$ and $S_j$ are neighboring Ising spins.
A. Bunde~\cite{bund74} has shown that in the correctly performed molecular field approximation
for ferro- and antiferromagnets
the correlation function $\langle S(\mathbf{q}) S(- \mathbf{q}) \rangle$ should fulfill the sum rule 
\begin{equation}
 N^{-1} \sum_{q}  \langle S(\mathbf{q}) S(- \mathbf{q}) \rangle = 1.
\end{equation}
The Ising model of the ferromagnet was considered~\cite{bund74} and the correlation 
function $\langle S(\mathbf{q}) S(- \mathbf{q}) \rangle$ was calculated
\begin{equation}
 \langle S(\mathbf{q}) S(- \mathbf{q}) \rangle = \Bigl [  N^{-1} \sum_{q} \frac{1}{1 - \beta J (\mathbf{q})}  \Bigr ]^{-1} \frac{1}{1 - \beta J (\mathbf{q})},
\end{equation}
which obviously fulfills the above sum rule. 
The Ising model and the Heisenberg model were  two the most explored models for the applications of the mean field theory.\\
It is of instruction to mention that earlier molecular-field concepts
described the mean-field in terms of some functional of
the average density of particles $\langle n \rangle$ (or, using the magnetic
terminology, the average magnetization $\langle M \rangle$), that
is, as $F[\langle n \rangle, \langle M \rangle]$. Using the modern language, one can
say that the interaction between the atomic spins $S_{i}$
and their neighbors can be equivalently described by
effective (or mean) field $ h^{(mf)}$. As a result one can write down
\begin{equation}
  M_{i} = \chi_{0} [ h_{i}^{(ext)} + h_{i}^{(mf)} ].
\end{equation}
The mean field $ h^{(mf)}$ can be represented in the form
(in the case $T >  T_{C}$)
\begin{equation}
  h^{(mf)} = \sum_{i} J(R_{ji})\langle S_{i} \rangle.
\end{equation}
Here, $ h^{ext}$ is the external magnetic field, $\chi_{0}$ is the system's
response function, and $ J(R_{ji})$ is the interaction
between the spins. In other words, in the mean-field
approximation a many-particle system is reduced to the
situation, where the magnetic moment at any site aligns
either parallel or anti-parallel to the overall magnetic
field, which is the sum of the applied external field and the molecular field. \\ 
Note that only the "\emph{averaged}" interaction
with $i$ neighboring sites was taken into account,
while the fluctuation effects were ignored. We see that the
mean-field approximation provides only a rough
description of the real situation and overestimates the
interaction between particles. Attempts to improve the
homogeneous mean-field approximation were undertaken
along different directions~\cite{kuz09,tyab,smart,rscom08,kuz78,kuznc94,kuzrnc02}. \\ An extremely
successful and quite nontrivial approach was developed
by L. Neel~\cite{kuz09,tyab,smart,rscom08}, who essentially formulated the concept
of \textbf{local mean fields} (1932). Neel assumed that the sign
of the mean-field could be both positive and negative.
Moreover, he showed that below some critical temperature
(the Neel temperature) the energetically most
favorable arrangement of atomic magnetic moments is
such, that there is an equal number of magnetic
moments aligned against each other. This novel magnetic
structure became known as the \textbf{antiferromagnetism}~\cite{tyab,kuz09}.\\
It was established that the antiferromagnetic interaction tends to align neighboring spins
against each other. In the one-dimensional case this corresponds
to an alternating structure, where an "\textbf{up}" spin
is followed by a "\textbf{down}" spin, and vice versa. Later it
was conjectured that the state made up from two
inserted into each other sublattices is the ground state of
the system (in the classical sense of this term). Moreover,
the mean-field sign there alternates in the "chessboard"  (staggered) order. \\ 
The question of the true antiferromagnetic
ground state is not completely clarified
up to the present time~\cite{kuz09,tyab,smart,rscom08,kuz78,kuznc94,kuzrnc02}. This is related to the
fact that, in contrast to ferromagnets, which have a
unique ground state, antiferromagnets can have several
different optimal states with the lowest energy. The
Neel ground state is understood as a possible form of
the system's wave function, describing the antiferromagnetic
ordering of all spins. Strictly speaking,
the ground state is the thermodynamically equilibrium
state of the system at zero temperature. Whether the
Neel state is the ground state in this strict sense or not,
is still unknown. It is clear though, that in the general
case, the Neel state is not an eigenstate of the Heisenberg
antiferromagnet's Hamiltonian. On the contrary,
similar to any other possible quantum state, it is only
some linear combination of the Hamiltonian eigenstates.
Therefore, the main problem requiring a rigorous
investigation is the question of Neel state 
stability~\cite{kuz09}. In some sense, only for infinitely large lattices,
the Neel state becomes the eigenstate of the Hamiltonian
and the ground state of the system. Nevertheless,
the sublattice structure is observed in experiments on
neutron scattering~\cite{kuz09}, and, despite certain worries,
the actual existence of sublattices is beyond doubt. \\
Once Neel's investigations were published, the
effective mean-field concept began to develop at a
much faster pace. An important generalization and
development of this concept was proposed in 1936 by
L. Onsager~\cite{onsag36} in the context of the polar liquid theory.
This approach is now called the \emph{Onsager reaction field} 
approximation. It became widely known, in particular,
in the physics of magnetic phenomena~\cite{wysin1,wysin2,medv07}.
In 1954, Kinoshita and Nambu~\cite{namb54} developed a systematic method
for description of many-particle systems in the framework
of an approach which corresponds to the \textbf{generalized mean-field} concept. N.D. Mermin~\cite{merm63} has 
analyzed the thermal Hartree-Fock
approximation~\cite{feld69} of Green function theory giving the free energy of a system not at zero temperature.\\
Kubo and Suzuki~\cite{suzuk68}    studied the applicability of the mean field approximation and showed that the
ordinary mean-field theory is restricted only to the region $k_{B}T \geq z J$, where $J$ denotes the strength of
typical interactions of the relevant system and $z$ the number of nearest neighbours.
Suzuki~\cite{suzuk86} has proposed a new type of fluctuating mean-field theory. In that approach the true critical
point $\tilde{T}_{C}$ differs from the mean-field value and the singularities of response functions are, in general, 
different from those of the Weiss mean-field theory~\cite{kuz09,smart}.\\
Lei Zhou and Ruibao Tao~\cite{tao97}  developed a complete Hartree-Fock mean-field method to study ferromagnetic systems 
at finite temperatures. With the help of the complete Bose transformation, they renormalized all the high-order 
interactions including both the dynamic and the kinetic ones based on an 
independent Bose representation, and obtained a set of compact self-consistent equations. Using their method, 
the spontaneous magnetization of an 
Ising model on a square lattice was investigated. The result is reasonably close to the exact one. Finally, they discussed 
the temperature dependences of the coercivities for magnetic systems and showed the hysteresis loops at 
different temperatures.\\
Later, various schemes of "effective mean-field theory taking into account correlations"
were proposed (see Refs.~\cite{kuz09,kuzrnc02}). We will see below  
that various mean-field approximations can be in principle
described in the framework of the variation principle  in terms of the Bogoliubov inequality~\cite{bogocp12,bb09,tyab,huber68,petqsm95}:
\begin{eqnarray}
\nonumber
 F =  - \beta^{-1} \ln (\textrm{Tr} e^{-\beta H } ) \leq \\ -
\beta^{-1} \ln (\textrm{Tr} e^{-\beta H_{\textrm{mod}}}) + \frac {\textrm{Tr} e^{ - \beta
H_{\textrm{mod}} } ( H - H_{\textrm{mod}})}{\textrm{Tr} e^{- \beta H_{\textrm{mod}} }}. \label{e83}
\end{eqnarray}
Here, $F$ is the free energy of the system under consideration,
whose calculation is extremely involved in the
general case. The quantity $H_{\textrm{mod}}$ is some trial Hamiltonian
describing the effective-field approximation. The
inequality (\ref{e83}) yields an upper bound for the free energy of a many-particle system.\\
It is well known, that the study of Hamiltonians describing strongly-correlated systems is an exceptionally
difficult many-particle problem, which requires
applications of various mathematical methods~\cite{kuz09,bach94,vbach97,lieb06,bach14a}.
In fact, with the exception of a few particular cases, even the ground state of the Hubbard model is
still unknown. Calculation of the corresponding quasiparticle spectra in the case of strong inter-electron correlations
and correct definition of the mean fields also turned out to be quite a complicated problem.\\
The Hamiltonian of the Hubbard model~\cite{kuz09} is given by:
\begin{equation}\label{e18}
H = \sum_{ij\sigma}t_{ij}a^{\dagger
}_{i\sigma}a_{j\sigma} + U/2\sum_{i\sigma}n_{i\sigma}n_{i-\sigma}.
\end{equation}
The above Hamiltonian includes the repulsion of the
single-site intra-atomic Coulomb $U$, and $t_{ij}$, the one-electron
hopping energy describing jumps from a $j$ site
to an $i$ site. As a consequence of correlations electrons
tend to "avoid one another". Their states are best modeled
by atom-like Wannier wave functions $[\phi({\vec r} -{\vec R_{j}})]$.
The Hubbard model's Hamiltonian can be characterized
by two main parameters: $U$, and the effective band
width of tightly bound electrons
$$\Delta = (N^{-1}\sum_{ij}\vert t_{ij}\vert^{2})^{1/2}.$$
The band energy of Bloch electrons $\epsilon(\vec k)$ is given by
$$\epsilon(\vec k)     = N^{-1}\sum_{\vec k}t_{ij} \exp[- i{\vec
k}({\vec R_{i}} -{\vec R_{j}}],$$
where $N$ is the total number of lattice sites. Variations
of the parameter $\gamma = \Delta/U$ allow one to study two interesting
limiting cases, the band regime ($\gamma \gg 1$) and the
atomic regime ($\gamma \rightarrow 0$).\\
There are many different approaches to construction
of generalized mean-field approximations; however, all
of them have a special-case character. The method of
irreducible Green  functions~\cite{kuz09,kuz78,kuznc94,kuzrnc02} allows one to tackle this
problem in a more systematic fashion.\\
The efficiency of the method of the irreducible
Green functions for description of normal and superconducting
properties of systems with a strong interaction
and complicated character of the electron spectrum
was demonstrated in the papers~\cite{kuz09,kuz78,kuznc94,kuzrnc02}. Let us
consider the Hubbard model (\ref{e18}). The properties of this
Hamiltonian are determined by the relationship
between the two parameters: the effective band's width $\Delta$
and the electron's repulsion energy $U$. Drastic transformations
of the metal-dielectric phase transition's
type take place in the system as the ratio of these
parameters changes. Note that, simultaneously, the
character of the system description must change as
well, that is, we always have to describe our system by
the set of relevant variables. In the case of weak correlation~\cite{kuz09,kuz78,kuznc94,kuzrnc02} the
corresponding set of relevant variables contains the
ordinary second-quantized Fermi operators and $a^{\dagger}_{i\sigma}$ ¨ $a_{i\sigma}$,
as well as the number of particles operator $n_{i\sigma} = a^{\dagger}_{i\sigma}a_{i\sigma}.$
In the case of strong correlation~\cite{kuz09,kuz78,kuznc94,kuzrnc02} the problem is highly complicated.\\
The Green function in the generalized mean-field's approximation has the following very 
complicated functional structure~\cite{kuz09,kuz78,kuznc94,kuzrnc02}:
\begin{equation}
\label{e125} G^{MF}_{k\sigma}( \omega) = \frac{\omega -
(n^{+}_{-\sigma}E_{-} + n^{-}_{-\sigma}E_{+}) - \lambda(k)}{
(\omega -E_{+} - n^{-}_{-\sigma}\lambda_{1}(k))(\omega - E_{-} -
 n^{+}_{-\sigma}
\lambda_{2}(k)) -
n^{-}_{-\sigma}n^{+}_{-\sigma}\lambda_{3}(k)\lambda_{4}(k)}.
\end{equation}
Here, the quantities $\lambda_{i}(k)$ are the components of the
generalized mean field, which cannot be reduced to the
functional of the mean particle's densities. The expression
for Green function (\ref{e125}) can be written down in the form of the
following \textbf{generalized two-pole solution}
\begin{eqnarray}
\label{e126} G^{MF}_{k\sigma}( \omega) = \frac
{n^{+}_{-\sigma}(1 + cb^{-1})}{a - db^{-1}c} + \frac
{n^{-}_{-\sigma}(1 + da^{-1})}{b - ca^{-1}d} \approx
\nonumber\\
\frac {n^{-}_{-\sigma}}{\omega - E_{-} -
n^{+}_{-\sigma}W^{-}_{k-\sigma}} + \frac {n^{+}_{-\sigma}}{\omega
- E_{+} - n^{-}_{-\sigma}W^{\dagger}_{k-\sigma}},
\end{eqnarray}
where
\begin{eqnarray}
\label{e127}
n^{+}_{-\sigma}n^{-}_{-\sigma}W^{\pm}_{k-\sigma} =
N^{-1}\sum_{ij} t_{ij}
\exp[-ik(R_{i} -R_{j})] \times  \\
\left ((\langle a^{\dagger}_{i-\sigma}n^{\pm}_{i\sigma}a_{j-\sigma} \rangle +
\langle a_{i-\sigma}
n^{\mp}_{i\sigma}a^{\dagger}_{j-\sigma} \rangle)    \right . + \nonumber \\
 \left . (\langle n^{\pm}_{j-\sigma}n^{\pm}_{i-\sigma} \rangle +
\langle a_{i\sigma}a^{\dagger}_{i-\sigma}
a_{j-\sigma}a^{\dagger}_{j\sigma} \rangle -
\langle a_{i\sigma}a_{i-\sigma}a^{\dagger}_{j-\sigma}
a^{\dagger}_{j\sigma} \rangle) \right). \nonumber
\end{eqnarray}
Green  function (\ref{e126}) is the \emph{most general solution}
of the Hubbard model within the generalized mean field 
approximation. Equation (\ref{e127}) is nothing else
but the explicit expression for the generalized mean field.
As we see, this mean field is not a functional of
the mean particle's densities. The solution (\ref{e126}) is
more general than the solution "\emph{Hubbard III}"~\cite{kuz09} and
other two-pole solutions.
Hence it was shown in the papers~\cite{kuz09,kuz78,kuznc94,kuzrnc02}  that the solution "\emph{Hubbard I}"~\cite{kuz09}
is a particular case of the solution (\ref{e126}), which
corresponds to the additional approximation
\begin{equation}
\label{e128} n^{+}_{-\sigma}n^{-}_{-\sigma}W^{\pm}(k) \approx
N^{-1}\sum_{ij}t_{ij} {\exp[-ik(R_{i} -
R_{j})]} \langle n^{\pm}_{j-\sigma}n^{\pm}_{i-\sigma}\rangle.
\end{equation}
Assuming $\langle n_{j-\sigma}n_{i-\sigma}\rangle \approx n^{2}_{-\sigma},$ we obtain the approximation
"\emph{Hubbard I}"~\cite{kuz09,kuz78,kuznc94,kuzrnc02}. Thus, we have shown that in
the cases of systems of strongly correlated particles
with a complicated character of quasiparticle spectrums
the generalized mean fields can have quite a nontrivial
structure, which is difficult to establish by using
any kind of independent considerations. The method of
irreducible Green functions allows one to obtain this structure in the
most general form.\\
One should note that
the BCS-Bogoliubov superconductivity theory~\cite{bogocp12,bb09,tyab,huber68,petqsm95}
is formulated in terms of a trial (approximating)
Hamiltonian  $H_{\textrm{mod}}$, which is a quadratic form with respect to
the second-quantized creation and annihilation operators,
including the terms responsible for anomalous (or
non-diagonal) averages. For the single-band Hubbard
model the BCS-Bogoliubov functional of generalized
mean fields can be written in the following form~\cite{vkp82,khp83,wyku83,kuzcmp10}
\begin{equation} \label{e84}
\Sigma^{c}_{\sigma} =  U 
\begin{pmatrix}   \langle a^{\dagger}_{i-\sigma}
a_{i-\sigma} \rangle & - \langle a_{i\sigma} a_{i-\sigma} \rangle \\
- \langle a^{\dagger}_{i-\sigma} a^{\dagger}_{i\sigma} \rangle &
- \langle a^{\dagger}_{i\sigma} a_{i\sigma} \rangle \\  
\end{pmatrix}. 
\end{equation}
The \emph{anomalous} (or nondiagonal) mean values in this
expression fix the vacuum state of the system exactly in
the BCS-Bogoliubov form. \\
It is worth mentioning that the modern microscopic theory of superconductivity was given a
rigorous mathematical formulation in the classic works of Bogoliubov and
co-workers~\cite{bogocp12,bb09,tyab,huber68,petqsm95} simultaneously with the Bardeen, Cooper, and Sehrieffer (BCS)  theory. It was shown that 
the equations of superconductivity can be derived from the fundamental electron-ion and electron-electron
interactions. The set of equations obtained is known as the Eliashberg
equations. They enable us to investigate the electronic and lattice
properties of a metal in both the normal and superconducting states.
Moreover, the Eliashberg equations are appropriate to the description of
strong coupling superconductors, in contrast to the  
equations, which are valid in the weak coupling regime and describe the
electron subsystem in the superconducting state only.\\
In paper~\cite{vkp82} on the basis of the BCS-Bogoliubov functional of generalized
mean fields a system of equations of superconductivity for the tight-binding electrons in the
transition metal described by the Hubbard Hamiltonian was derived. The electron-phonon
interaction was written down for the "rigid ion model". Neglecting the vertex corrections
in the self-energy operator the closed system of equations was obtained.\\
In paper~\cite{khp83} this approach was extended for the Barisic-Labbe-Friedel model of a transition metal.
The renormalized electron and phonon spectra of the model were derived 
using the method of irreducible Green  functions~\cite{kuz09,kuz78,kuznc94,kuzrnc02} in a  the self-consistent way. 
For the band and atomic limits of the Hubbard model the explicit solutions for the electron
and phonon energies were obtained.  The energy gap, appearing between electron
bands in the strong correlation limit, persists in that calculations. The Eliashberg-type equations of 
superconductivity were obtained also.\\
The equations of strong coupling superconductivity in disordered transition
metal alloys have been derived in paper~\cite{wyku83}  by means of  irreducible  Green  functions method
and on the basis of the alloy version of the Barisic-Labbe-Friedel model
for electron-ion interaction. The configurational averaging has been
performed by means of the coherent potential approximation. Making
some approximations,  the formulas for the superconducting transition temperature
$T_{C}$ and the electron-phonon coupling constant  have been obtained. These depend on
the alloy component and total densities of states, the phonon Green  function,
and the parameters of the model.\\ 
To summarize, various schemes of "effective mean-field theory"  taking into account correlations
were proposed~\cite{kuz78,kuznc94,kuzrnc02,liu67,petpl68,peter68,marsh69,dmatt79,kane80,fah81,kat83,bali85,howl03,kris11}.
The main efforts were directed to the aim to describe  suitably
the collective behavior of particles in terms of effective-field distribution which
satisfies a self-consistent condition. However, although the self-consistent field approximation often is 
a reasonable approximation away from the critical point, it usually breaks down in its immediate
neighborhood.\\ 
It is of importance to stress again that from our point of view,   
in real mean-field theory, the mean field appearing in the single-site problem should be  a scalar or 
vectorial \textbf{time-independent} quantity. 
%
%
%
%
%
\section{Symmetry Broken Solutions} 
%
The formalism of the previous sections may be extended to incorporate the broken symmetry solutions~\cite{kuz09,kuz10,matt68}
of the interacting many particle systems, e.g. the pairing effects present in superconductors~\cite{bb09,bb14,petqsm95}, etc.
Our purpose in this section is to attract attention to subtle points which are essential for establishing a connection
of the generalized mean field  approximation and the broken symmetry solutions~\cite{kuz09,kuz10,matt68}.\\
It is well known that a symmetry can be exact or approximate. 
Symmetries inherent in the physical laws may be dynamically and spontaneously broken, i.e., they may not
manifest themselves in the actual phenomena. It can be as well broken by certain reasons~\cite{namb07,fuji11}. \\
Within the literature the 
term \emph{broken symmetry} is used both very often and with different meanings. There are two terms, the 
spontaneous breakdown of symmetries and dynamical symmetry breaking, which sometimes have been used as opposed 
but such a distinction is irrelevant. However, the two terms may be used interchangeably. It should be stressed that
a symmetry implies degeneracy. In general there are a multiplets of equivalent states related to each other by congruence
operations. They can be distinguished only relative to a weakly coupled external environment which breaks the
symmetry. Local gauged symmetries, however, cannot be broken this way because such an extended environment is not
allowed (a superselection rule), so all states are singlets, i.e., the multiplicities are not observable except possibly for
their global part. \\
It is known that when the Hamiltonian of a system is invariant under a symmetry operation, but the
ground state is not, the symmetry of the system can be spontaneously broken. 
Symmetry breaking is termed \emph{spontaneous} when there is no explicit term in a Lagrangian which
manifestly breaks the symmetry.\\
Peierls~\cite{pei91,pei92} gave  a general definition of the notion of the spontaneous breakdown of symmetries which is
suited equally well for the physics of particles and condensed matter physics.   According to Peierls~\cite{pei91,pei92}, the term \emph{broken symmetries}
relates to situations in which symmetries which we expect to hold are valid only approximately or fail completely 
in certain situations.\\
The intriguing mechanism of spontaneous symmetry breaking is a unifying concept that lie at the basis of most 
of the recent developments in theoretical physics, from statistical mechanics to many-body theory and to elementary 
particles theory~\cite{namb07,fuji11}. 
The existence of degeneracy in the energy states of a quantal system is related to the invariance 
or symmetry properties of the system. By applying the 
symmetry operation to the ground state, one can transform it to a different but equivalent ground state. Thus the
ground state is degenerate, and in the case of a continuous symmetry, infinitely degenerate. The real, or relevant,
ground state of the system can  only be one of these degenerate states. A system may exhibit the full symmetry of its
Lagrangian, but it is characteristic of infinitely large systems that they also may condense into states 
of lower symmetry.\\
It should be pointed out that  Bogoliubov's method of quasiaverages~\cite{bb09,bb14,petqsm95}  gives the deep foundation and clarification of the concept of broken symmetry. It makes the emphasis on the notion of   
degeneracy and plays an important role in equilibrium statistical mechanics of many-particle systems. According to that concept, infinitely small 
perturbations can trigger macroscopic responses in the system if they break some symmetry and remove the related degeneracy (or quasi-degeneracy) 
of the equilibrium state. As a result, they can produce macroscopic effects even when the perturbation magnitude is tend to zero, provided that 
happens after passing to the thermodynamic limit~\cite{kuz14}. This approach has penetrated, directly or indirectly, many areas of 
the contemporary physics. \\
The article~\cite{kuz10} examines  the Bogoliubov's notion of quasiaverages, from the original papers~\cite{bb14}, through to modern theoretical concepts and ideas of how 
to describe both the degeneracy, broken symmetry and the diversity of the energy scales in the many-particle interacting systems. Current trends 
for extending and using Bogoliubov's ideas to quantum field theory and condensed matter physics problems were discussed, including microscopic 
theory of superfluidity and superconductivity, quantum theory of magnetism of complex materials, Bose-Einstein condensation, chirality of 
molecules, etc. 
Practical techniques covered include quasiaverages, Bogoliubov theorem on the singularity of $1/q^2$, Bogoliubov's inequality, and its applications 
to condensed matter physics.\\
It was demonstrated there that the profound and innovative idea of quasiaverages formulated by N.N. Bogoliubov, gives the so-called macro-objectivation of the 
degeneracy in domain of quantum statistical mechanics, quantum field theory and in the quantum physics in general.  \\
The quasiaverages may be obtained from the ordinary averages by using the cluster property which was formulated
by Bogoliubov~\cite{bb09,bb14,petqsm95}. This was first done when deriving the Boltzmann equations from the chain of equations
for distribution functions, and in the investigation of the model Hamiltonian in the theory 
of superconductivity~\cite{bb09,bb14,petqsm95}.  To demonstrate this
let us   consider averages (quasiaverages) of the form
\begin{equation}
\label{q11}
F(t_{1},x_{1}, \ldots t_{n},x_{n}) = \langle \ldots \Psi^{\dag}(t_{1},x_{1}) \ldots \Psi(t_{j},x_{j}) \ldots \rangle,
\end{equation}
where the number of creation operators $\Psi^{\dag}$ may be not equal to the number of annihilation operators $\Psi$. We 
fix times and split the arguments $(t_{1},x_{1}, \ldots t_{n},x_{n})$ into several 
clusters $( \ldots , t_{\alpha},x_{\alpha}, \ldots ), \ldots ,$
$( \ldots , t_{\beta},x_{\beta}, \ldots ).$  Then it is reasonably to assume that the
distances between all clusters $|x_{\alpha} - x_{\beta}|$ tend to infinity. Then, according to the cluster property, the
average value (\ref{q11}) tends to the product of averages of collections of operators with the arguments
$( \ldots , t_{\alpha},x_{\alpha}, \ldots ), \ldots ,$  $( \ldots , t_{\beta},x_{\beta}, \ldots )$ 
\begin{equation}
\label{q12}
\lim_{|x_{\alpha} - x_{\beta}| \rightarrow \infty}  F(t_{1},x_{1}, \ldots t_{n},x_{n}) = 
F( \ldots , t_{\alpha},x_{\alpha}, \ldots )  \ldots   F( \ldots , t_{\beta},x_{\beta}, \ldots ).
\end{equation}
For equilibrium states with small densities and short-range potential, the validity of this property can 
be proved~\cite{bb09,bb14,petqsm95}.
For the general case, the validity of the cluster property has not yet been proved. Bogoliubov
formulated it not only for ordinary averages but also for quasiaverages, i.e., for anomalous averages, too. It works
for many important models, including the models of superfluidity and superconductivity~\cite{bb09,bb14,petqsm95}.\\
In his work  \emph{The Theory of Superfluidity}~\cite{nnb47},  Bogoliubov gave a microscopic explanation of
the phenomenon of superfluidity~\cite{nbog94,grif09}. Before his works, there were the phenomenological theories 
which were based on an assumption about the form of the spectrum of elementary excitations.
Bogoliubov has started from the general Hamiltonian for Bose systems and assumed that a macroscopic number
of particles are found in the ground state with zero momentum, and therefore the creation and
annihilation operators of particles with zero momentum are $c$-numbers~\cite{sank10}. As a result a definite \emph{approximating}
Hamiltonian was obtained, consisting from a quadratic form of the creation and annihilation operators. The
usual perturbation theory proved to be inapplicable to it because of the strong interaction of
particles with opposite momenta. Therefore the Hamiltonian was diagonalized with the help of
the canonical transformations (the Bogolyubov $u-v$ transformations). This permitted one to calculate 
the spectrum of elementary perturbations outside the framework of perturbation theory.
Decomposing the field operators into $c$-numerical and operator parts, Bogoliubov in fact
introduced into quantum theory the method of \emph{spontaneous symmetry breakdown} for systems with
degenerate ground state. This method was rediscovered in quantum field theory a decade later~\cite{kuz10}.\\
To illustrate  these statements consider Bogoliubov's theory of a Bose-system with separated condensate, which is given by the 
Hamiltonian~\cite{bb09,bb14,petqsm95}
\begin{eqnarray}\label{q13}
H_{\Lambda}  = \int_{\Lambda}\Psi^{\dag}(x)(- \frac{\Delta}{2m})\Psi(x)dx - \mu \int_{\Lambda}\Psi^{\dag}(x)\Psi(x)dx \\ \nonumber
+ \frac{1}{2} \int_{\Lambda^{2}}\Psi^{\dag}(x_{1}) \Psi^{\dag}(x_{2}) \Phi (x_{1} - x_{2})   \Psi(x_{2})\Psi(x_{1})dx_{1} dx_{2}.
\end{eqnarray} 
This Hamiltonian can be written also in the following form
\begin{eqnarray}\label{q14}
H_{\Lambda}  =  H_{0} + H_{1} = \int_{\Lambda}\Psi^{\dag}(q)(- \frac{\Delta}{2m})\Psi(q)d q \\ \nonumber
+ \frac{1}{2} \int_{\Lambda^{2}}\Psi^{\dag}(q) \Psi^{\dag}(q') \Phi (q - q')   \Psi(q')\Psi(q)d q d q'.
\end{eqnarray} 
Here, $\Psi(q)$, and $\Psi^{\dag}(q)$  are the operators of annihilation and creation of bosons. They satisfy the 
canonical commutation relations
\begin{equation}
\label{q15}
 [\Psi(q),\Psi^{\dag}(q')]  = \delta (q - q'); \quad [\Psi(q),\Psi(q')]  = [\Psi^{\dag}(q),\Psi^{\dag}(q')]  = 0.
\end{equation}
The system of bosons is contained in the cube $A$ with the edge $L$ and volume $V$. It was assumed that it satisfies
periodic boundary conditions and the potential $\Phi(q)$ is spherically symmetric and proportional 
to the small parameter.
It was also assumed that, at temperature zero, a certain macroscopic number of particles having a nonzero
density is situated in the state with momentum zero.\\
The operators $\Psi(q)$, and $\Psi^{\dag}(q)$  are represented in the form
\begin{equation}
\label{q16}
  \Psi(q)  = a_{0}/\sqrt{V};  \quad \Psi^{\dag}(q) = a^{\dag}_{0}/\sqrt{V},
\end{equation}
where $a_{0}$ and $a^{\dag}_{0}$ are the operators of annihilation and creation of particles with momentum zero.\\
To explain the phenomenon of
superfluidity~\cite{bb14,nnb47}, one should calculate the spectrum of the Hamiltonian, which is quite a difficult problem. Bogoliubov
suggested the idea of approximate calculation of the spectrum of the ground state and its elementary excitations
based on the physical nature of superfluidity. His idea consists of a few assumptions. The main assumption
 is   that at temperature zero the macroscopic number of particles (with nonzero density) has the
momentum zero. Therefore, in the thermodynamic limit~\cite{kuz14}, the operators $a_{0}/\sqrt{V}$ and $a^{\dag}_{0}/\sqrt{V}$ commute
\begin{equation}
\label{q17}
\lim_{V \rightarrow \infty} \left [ a_{0}/\sqrt{V}, a^{\dag}_{0}/\sqrt{V} \right ] = \frac{1}{V} \rightarrow 0
\end{equation}
and are $c$-numbers. Hence, the operator of the number of particles $N_{0} =a^{\dag}_{0}a_{0}$  is a $c$-number, too.\\
D. Ya. Petrina~\cite{petri96} shed an additional light on the problem of 
an approximation of general Hamiltonians by Hamiltonians of the theories of superconductivity and superfluidity.
In his highly interesting paper~\cite{petri96} Petrina pointed out that
the model Hamiltonian of the theory of superconductivity~\cite{bb09,petqsm95} can be obtained from the general Hamiltonian 
for Fermi systems if the Kronecker symbol, which expresses the law of conservation of momentum in the interaction 
Hamiltonian, is replaced by two Kronecker symbols so that only particles with opposite momenta interact. 
The model Hamiltonian of the theory of superfluidity can be obtained from the 
general Hamiltonian for Bose systems if we replace the Kronecker symbol, which expresses the law of conservation of 
momentum, by several Kronecker symbols, preserving only the terms that contain at least two operators with momenta zero 
in the interaction Hamiltonian. This list of model systems can be continued~\cite{stoof97}.\\
The concept of quasiaverages was introduced by Bogoliubov on the basis of an analysis of many-particle systems with
a degenerate statistical equilibrium state. Such states are inherent to various physical many-particle systems. Those are
liquid helium in the superfluid phase, metals in the superconducting state, magnets in the ferromagnetically ordered state,
liquid crystal states, the states of superfluid nuclear matter, etc. \\
In many-body interacting systems, the symmetry is important in classifying
different phases and in understanding  the phase transitions
between them. According to Bogoliubov's ideas~\cite{bb09,petqsm95,bb14,nnb47,matt68} 
in each condensed phase, in addition to the normal process, there is an
anomalous process (or processes) which can take place because of
the long-range internal field, with a corresponding propagator.
Additionally, the Goldstone theorem~\cite{kuz10} states that, in a
system in which a continuous symmetry is broken ( i.e.  a
system such that the ground state is not invariant under the
operations of a continuous unitary group whose generators commute
with the Hamiltonian), there exists a collective mode with
frequency vanishing, as the momentum goes to zero. For
many-particle systems on a lattice, this statement needs a proper
adaptation.   In the above form, the Goldstone theorem is true
only if the condensed and  normal phases have the same
translational properties. When translational symmetry is also
broken, the Goldstone mode appears at a zero frequency but at
nonzero momentum, e.g., a crystal and a helical
spin-density-wave  ordering (see for discussion Refs.~\cite{kuz09,kuz99}). \\
The antiferromagnetic state is
characterized by a spatially changing component of magnetization which
varies in such a way that the net magnetization of the system is zero.
The concept of antiferromagnetism of localized spins which is based on
the Heisenberg model and the two-sublattice Neel ground state is
relatively well founded contrary to the antiferromagnetism of
delocalized or itinerant electrons. The itinerant-electron picture is
the alternative conceptual picture for magnetism~\cite{her22}.
In the antiferromagnetic many-body
problem there is an additional "symmetry broken" aspect~\cite{kuz09,kuz99}.  For an
antiferromagnet, contrary to ferromagnet, the one-electron Hartree-Fock
potential can violate the translational crystal symmetry. The period of
the antiferromagnetic spin structure $L$ is greater than the lattice
constant $a$. The Hartree-Fock is the simplest approximation but neglects
the important dynamical part. To include the dynamics one should take
into consideration the correlation effects.\\
The anomalous propagators for an interacting many-fermion system
corresponding to the ferromagnetic (FM) , antiferromagnetic
(AFM),  and superconducting (SC) long-range ordering are given by
\begin{eqnarray} \label{eq.71}
FM: G_{fm} \sim \langle \langle a_{k\sigma};a^{\dagger}_{k-\sigma} \rangle \rangle,\\
\nonumber AFM: G_{afm} \sim \langle \langle a_{k+Q\sigma};a^{\dagger}_{k+Q'\sigma'} \rangle \rangle,\\
\nonumber SC: G_{sc} \sim \langle \langle a_{k\sigma};a_{ - k -\sigma} \rangle \rangle.\\
\nonumber
\end{eqnarray} 
In the spin-density-wave case, a particle picks up a momentum $Q - Q'$
from scattering against the periodic structure of the spiral (nonuniform) internal field, and has its spin changed from
$\sigma$ to $\sigma'$ by the spin-aligning character of the
internal field.  The Long-Range-Order (LRO) parameters are:
\begin{eqnarray} \label{eq.72}
FM: m = 1/N\sum_{k\sigma} \langle a^{\dagger}_{k\sigma}a_{k-\sigma} \rangle,\\
\nonumber AFM:
M_{Q} = \sum_{k\sigma} \langle a^{\dagger}_{k\sigma}a_{k+Q-\sigma} \rangle, \\
\nonumber SC: \Delta = \sum_{k} \langle a^{\dagger}_{-k\downarrow}a^{\dagger}_{k \uparrow} \rangle.\\ \nonumber
\end{eqnarray} 
It is of
importance to note that the long-range order parameters are
functions of the internal field, which is itself a function of the
order parameter. There is a more mathematical way of formulating
this assertion. As it was stressed  earlier~\cite{kuz10}, the notion
 \emph{symmetry breaking}  means that the state fails to have the
symmetry that the Hamiltonian has.\\
In terms of the theory of quasiaverages, a true breaking of symmetry can arise only if there are
infinitesimal "source fields".   Indeed, for the rotationally and
translationally invariant Hamiltonian,  suitable source terms
should be added:
\begin{eqnarray} \label{eq.73}
FM:  \varepsilon\mu_{B}
H_{x}\sum_{k\sigma}a^{\dagger}_{k\sigma}a_{k-\sigma},\\ \nonumber
AFM:
\varepsilon \mu_{B} H \sum_{kQ} a^{\dagger}_{k\sigma}a_{k+Q-\sigma},\\
\nonumber SC: \varepsilon v \sum_{k} (a^{\dagger}_{- k \downarrow}
a^{\dagger}_{k \uparrow} + a_{k \uparrow} a_{ - k \downarrow}),
\end{eqnarray} 
where $\varepsilon \rightarrow 0$ is to be taken at
the end of calculations.\\ For example,  broken symmetry
solutions of the spin-density-wave type  imply that the vector $Q$ is a measure
of the inhomogeneity or breaking of translational symmetry. \\ In this context the
Hubbard model is a very interesting tool for  analyzing   the
broken symmetry  concept~\cite{kuz78,kuznc94,kuzrnc02}. It is possible to show that
antiferromagnetic state and more complicated states (e.g. 
ferrimagnetic) can be made eigenfunctions of the self-consistent
field equations within an "extended" (or generalized) mean-field approach,
assuming that the  \emph{anomalous}  averages
$\langle a^{\dagger}_{i\sigma}a_{i-\sigma} \rangle$ determine the behaviour of
the system on the same footing as the "normal" density of
quasiparticles $\langle a^{\dagger}_{i\sigma}a_{i\sigma} \rangle$.  It is
clear, however, that these ``spin-flip" terms break the rotational
symmetry of the Hubbard Hamiltonian. For the single-band Hubbard
Hamiltonian, the averages $\langle a^{\dagger}_{i-\sigma}a_{i, \sigma} \rangle
= 0$ because of the rotational symmetry of the Hubbard model.  
The
inclusion of   \emph{anomalous}  averages leads to the following  approximation
\begin{equation}  
n_{i-\sigma}a_{i\sigma} \approx \langle n_{i-\sigma} \rangle a_{i\sigma} -
\langle a^{\dagger}_{i-\sigma}a_{i\sigma} \rangle a_{i-\sigma}.
\end{equation}
Thus, in addition to the standard Hartree-Fock term, the new
 so-called ``\emph{spin-flip}" terms are retained~\cite{kuz99}. This example
clearly shows that the structure of  mean field follows from the
specificity of the problem and should be defined in a proper
way.  So, one needs a properly defined effective Hamiltonian
$H_{\rm eff}$.  In paper~\cite{kuz99} we thoroughly analyzed
 the proper definition of the irreducible Green functions which
includes the ``spin-flip" terms for the case of itinerant
antiferromagnetism of correlated lattice fermions. For
the single-orbital Hubbard model~\cite{kuz78,kuznc94,kuzrnc02,kuz99}, the definition of the
"irreducible" part should be modified in the following way:
\begin{eqnarray} \label{eq.75}
^{(ir)}\langle \langle a_{k+p\sigma}a^{\dagger}_{p+q-\sigma}a_{q-\sigma} \vert
a^{\dagger}_{k\sigma} \rangle \rangle_ {\omega} =
\langle \langle a_{k+p\sigma}a^{\dagger}_{p+q-\sigma}a_{q-\sigma}\vert
a^{\dagger}_{k\sigma} \rangle \rangle_{\omega} - \nonumber\\ \delta_{p,
0}\langle n_{q-\sigma} \rangle G_{k\sigma} -
\langle a_{k+p\sigma}a^{\dagger}_{p+q-\sigma} \rangle \langle \langle a_{q-\sigma} \vert
a^{\dagger}_{k\sigma} \rangle \rangle_{\omega}.
\end{eqnarray}
From this definition it follows that this way of introduction of
the irreducible Green functions broadens the initial algebra of  operators and the
initial set of the Green functions.  This means that the ``actual" algebra of
 operators must include the spin-flip terms from the beginning,
namely:  $(a_{i\sigma}$, $a^{\dagger}_{i\sigma}$, $n_{i\sigma}$,
$a^{\dagger}_{i\sigma}a_{i-\sigma})$. The corresponding initial
Green function will be of the form
\begin{equation}
 \begin{pmatrix}
 \langle \langle a_{i\sigma}\vert a^{\dagger}_{j\sigma} \rangle \rangle & \langle \langle a_{i\sigma}\vert a^{\dagger}_{j-\sigma} \rangle \rangle \\
\langle \langle a_{i-\sigma}\vert a^{\dagger}_{j\sigma} \rangle \rangle & \langle \langle a_{i-\sigma}\vert a^{\dagger}_{j-\sigma} \rangle \rangle\\
\end{pmatrix}.
\end{equation}
With this definition, one
introduces the so-called anomalous (off-diagonal) Green functions which fix
the relevant vacuum and select the proper symmetry broken
solutions. In fact, this approximation was
 investigated earlier by Kishore and Joshi~\cite{kisor71}. They
clearly pointed out that they assumed a system to be magnetized in
the $x$ direction instead of the conventional $z$ axis.  \\ The
problem of finding  the superconducting, ferromagnetic and antiferromagnetic
"symmetry broken" solutions of the correlated lattice fermion
models within irreducible Green functions method was investigated in Refs.~\cite{kuz09,kuz78,kuznc94,kuzrnc02,kuz99}. A
unified scheme for the construction of \textbf{generalized mean fields} (elastic scattering corrections) 
and self-energy (inelastic scattering) in terms of the Dyson equation was  generalized in
order to include  the "source fields". The "symmetry broken"
dynamic solutions of the Hubbard model  which correspond to
various types of itinerant antiferromagnetism were  discussed as well~\cite{kuz09,kuz78,kuznc94,kuzrnc02,kuz99}.
This approach complements previous studies of microscopic theory
of the Heisenberg antiferromagnet~\cite{kuzmar90} and clarifies the
 concepts of Neel sublattices for localized and
itinerant antiferromagnetism and "spin-aligning fields" of
correlated lattice fermions.\\
We shall see shortly  that in order to discuss  the mean field theory  (and generalized mean fields) on the firm ground 
the Bogoliubov inequality provides the formal basis and effective general approach.
%
%
%
%
%
%
\section{The Mathematical Tools}  
%
Before entering fully into our subject, we must recall some basic statements. This will be necessary
for the following discussion.\\
The number of inequalities in mathematical physics is extraordinary plentiful and the literature on inequalities 
is vast~\cite{bulle98,bulle03,lieb03,carl09,gold65,thom65,arlieb70,rusk72,arak73,elieb76,rusk90,algebr10,bik11}.   
The physicists are interested mostly in intuitive, physical forms of inequalities rather than in their most general versions. 
Often it is easier to catch the beauty and importance of original versions rather than decoding their later, abstract forms.\\
Many inequalities are of a great use and directly related with the notion of entropy, especially with
quantum entropy~\cite{carl09,petz04}. 
The von Neuman entropy of $\rho$ $\in  \mathbf{S_{n}}$,  $S(\rho)$, is defined by
\begin{equation}
   S(\rho) =  - \textrm{Tr}(\rho \log \rho) .
\end{equation}
The operator $\rho \log \rho$  is defined using the spectral theorem~\cite{carl09}.
Here $\mathbf{S_{n}}$ denotes the set of density matrices $\rho$  on $\mathbb{C}^{n}$.\\
In fact, $S(\rho)$ depends on $\rho$ only through its eigenvalues. 
\begin{equation}
 S(\rho) = - \sum_{j =1}^{n} \lambda_{j} \log \lambda_{j}.
\end{equation}
Otherwise put, the von Neumann entropy is unitarily invariant; i.e.,
\begin{equation}
 S(U \rho U^{*})  = S(\rho).
\end{equation}
The convexity condition leads to~\cite{carl09}
\begin{equation}
 - S(\rho) = - \log (n).
\end{equation}
This equality valid if and only if each  $\lambda_{j} = 1/n $  Thus, one may arrive at~\cite{carl09}
\begin{equation}\label{ineqmt}
0 \leq S(\rho)  \leq \log n 
\end{equation}
for all $\rho$ $\in  \mathbf{S_{n}}$, and there is equality on the left if  $\rho$ is a pure state, and there is
equality on the right if  $\rho  = (1/n)I$.
Actually, $S(\rho)$  is not only a strictly concave function of the eigenvalues of $\rho$, it
is strictly concave function of $\rho$ itself.\\
The notions of convexity and concavity of trace functions~\cite{carl09} are of great importance in   mathematical 
physics~\cite{ehlieb73,tropp12}. Inequalities for quantum mechanical entropies and related concave trace functions
play a fundamental role in quantum information theory as well~\cite{carl09,petz04}.\\
A function $f$ is \emph{convex} in a given interval if its second derivative is
always of the same sign in that interval. The sign of the second derivative can
be chosen as positive (by multiplying by $(- 1)$ if necessary).
Indeed,  the  notion of convexity means that if $d^{2} f/d x^{2} > 0$
in a given interval, $x_{j}$ are a set of points in that interval, $p_{j}$
are a set of weights such that $p_{j} \geq 0$, which have the property $\sum_{j} p_{j} = 1$, then
\begin{equation}
 \sum_{j} p_{j} f(x_{j}) \geq f  \Bigl (\sum_{j} p_{j} x_{j} \Bigr ).
\end{equation}
The equality will be valid only if $x_{j} = \langle x \rangle = \sum_{j} p_{j} x_{j}$.
In other words, a real-valued function $f(x)$ defined on an interval is called convex (or convex downward or concave upward)
if the line segment between any two points on the graph of the function lies \textbf{above} the graph, 
in a Euclidean space (or more generally a vector space) of at least two dimensions. Equivalently, a function 
is \emph{convex} if its epigraph (the set of points on or above the graph of the function) is a convex set.\\ 
A real-valued function $f$ on an interval (or, more generally, a convex set in vector space) is said to be concave if, 
for any $x_{1}$ and $x_{2}$ in the interval and for any $\alpha$ in $[0,1]$,
\begin{equation}
  f\Bigl ((1-\alpha)x_{1} +(\alpha)x_{2} \Bigr )  \geq (1 - \alpha) f(x_{1}) + (\alpha) f(x_{2}).
\end{equation}
A function $f(x)$ is \emph{concave} over a convex set if and only if the function $- f(x)$ is a convex function over the set.\\
As an example we mentioned above briefly   a reason this concavity matters, pointing to the
inequality (\ref{ineqmt})   that was   deduced from the concavity of the entropy as a
function of the eigenvalues of $\rho$.\\
It is of importance to stress that in quantum statistical mechanics, equilibrium states are determined by maximum
entropy principles~\cite{carl09}, and the fact that
\begin{equation}
\sup  S(\rho) \Bigr|_{\rho \in  \mathbf{S_{n}}} = \log n,
\end{equation}
reflects the famous Boltzmann  formulae
\begin{equation}
 S = k_{B} \log W.
\end{equation}
It follows from Boltzmann definition
that the entropy is larger if $\rho$ is smeared out, where $\rho$ is the probability density on phase space.
The microscopic definition of entropy given
by Boltzmann does not, by itself, explain the second law of thermodynamics.
even in  classical physics. The task to formulate
these questions in a quantum framework    was addressed by
Oskar Klein in his seminal paper~\cite{klein32} of 1931. He found a fundamentally new
way for information to be lost hence entropy to increase, special to quantum
mechanics. This result was called Klein lemma~\cite{tropp12,rusk02,klein32}.\\
M. B. Ruskai~\cite{rusk02} has reviewed many fundamental properties of the quantum entropy~\cite{petz04} including
one important class of inequalities relates the entropy of subsystems to that of a composite
system.
That article presented self-contained proofs of the strong subadditivity inequality for
von Neumann  quantum entropy, $S(\rho)$, and some related inequalities for the quantum
relative entropy, most notably its convexity and its monotonicity under stochastic
maps. 
The approach to subadditivity and relative entropy presented was used to obtain conditions for equality in
properties of relative entropy, including its joint convexity and monotonicity.
In addition, the Klein   inequality was presented there in detail.\\
Indeed, the fact that the relative entropy is positive~\cite{rusk02}, i.e., $H(\rho_{1} ,\rho_{2})\geq 0 $ 
when  $ \textrm{Tr} \rho_{1}  = \textrm{Tr} \rho_{2}$, 
is an immediate consequence of the following fundamental convexity result due to Klein~\cite{klein32,aweh78,niels00}.   
The corresponding theorem~\cite{rusk02} states that for $A, B > 0$
\begin{equation}
 \textrm{Tr} \, A (\log A  -  \log B) \geq \textrm{Tr} (A - B),
\end{equation}
with equality if and only if $(A = B)$. \\ In more general form~\cite{carl09} the Klein   inequality may be formulated in
the following way.\\
For all $A, B \in  \mathbf{H}_{n}$, and all differentiable
convex functions $f: \mathbb{R} \to \mathbb{R}$, or for all $A, B \in  \mathbf{H}^{\dag}_{n}$
and all differentiable convex functions $f:  (0, \infty)  \to \mathbb{R}$
\begin{equation}\label{klemm}
  \textrm{Tr} \,   \Bigl ( f(A) - f(B) - (A - B)f'(B)  \Bigr )  \geq 0.
\end{equation}
In either case, if $f$ is strictly convex, there is equality if and only if $A = B$.\\
A few more words about Oskar Klein and his inequality will not be out of place here.
Oskar Klein (1894 - 1977) was famous Swedish theoretical physicist~\cite{} who worked on a wide variety of subjects~\cite{dani95}.
For example, the Klein-Gordon equation was the first relativistic wave equation. Oskar Klein was also a collaborator 
of Niels Bohr in Copenhagen.
It is interesting to note that Oskar Klein defended his thesis and was awarded his doctoral degree in 1921 for 
his work in physical chemistry about strong electrolytes.
In 1931 Oskar Klein~\cite{klein32,aweh78,niels00,dani95}, using his experience in both quantum and 
statistical mechanics, succeeded in solving the problem
 of whether the quantum statistics on molecular level can explain how the entropy increases with
time in accordance with the second law of thermodynamics. The problem in classical statistical mechanics had been
already noticed by Gibbs earlier. Klein proof~\cite{klein32,aweh78,niels00}, that used the statement that only the 
diagonal elements in the density matrix for the phase space of the particles are relevant for the entropy, 
has led him to the  Klein lemma. With Klein lemma, the entropy can increase according to the formula of   Boltzmann    microscopic definition, where it is
described with the number of states in the phase space. A useful and informative discussion of the Klein paper and Klein lemma was carried out in the book of R. Jancel~\cite{janc69}.\\ 
According to M. B. Ruskai~\cite{rusk02},  the closely related Peierls-Bogoliubov inequality  is sometimes used 
instead of Klein inequality.  Golden-Thompson and Peierls-Bogoliubov inequalities were extended
to von Neumann algebras, which have traces, by Ruskai~\cite{rusk72} (see also Ref.~\cite{zhao14}).   H. Araki~\cite{arak73} 
extended them to a general von Neumann algebra. This kind of investigations is particularly valuable since
the Bogoliubov inequality is remarkable because it  have significant applications in
statistical quantum mechanics~\cite{bb09,huber68,huber73,mazo74,bebian04,furu06}. It provides insight into a number of other interesting questions as well.\\
It will be of use to write down the mathematical formulation of Peierls-Bogoliubov inequalities which was provided 
by Carlen~\cite{carl09}.\\
Let us consider $A$ $\in  \mathbf{H_{n}}$,  and let $f$ be any convex function on  $\mathbb{R}$.
Let $\{u_{1}, \ldots, u_{n} \}$ be any orthonormal base of $\mathbb{C}^{n}$. Then
\begin{equation}
 \sum_{j =1}^{n}  \Bigl ( \langle u_{j}, A u_{j} \rangle \Bigr ) \leq \textrm{Tr} [f(A) ].
\end{equation}
There is equality if each $u_{j}$ is an eigenvector of $A$, and if $f$ is strictly convex, only
in this case.\\
Now consider the formulation of the generalized Peierls-Bogoliubov inequality~\cite{carl09}.
For every natural number $n$, the map $A \mapsto \log \Bigl ( \textrm{Tr} [\exp(A)] \Bigr ) $ is convex on $\mathbf{H_{n}}$.
As a consequence one may deduce~\cite{carl09} that
\begin{equation}
\log \Bigl ( \frac{\textrm{Tr} [\exp(A + B)]}{\textrm{Tr} [\exp(A)]} \Bigr ) \geq \frac{\textrm{Tr} [B \exp(A )]}{\textrm{Tr} [\exp(A)]}
\end{equation}
Frequently this relation, which has many uses, is referred to as the Peierls-Bogoliubov inequality.\\
It is worth noting that according to tradition the term    \emph{Gibbs-Bogoliubov inequality}~\cite{ishi68} is used 
for a classical statistical mechanical systems and term \emph{Peierls-Bogoliubov inequality}~\cite{carl09} for quantum statistical mechanical systems.\\
At the very least, it must have been meant to indicate that Peierls inequality does not have a classical analog, whereas 
Bogoliubov inequality has.
%
%
%
\section{Variational Principle of Bogoliubov}  
%
%
It is known that there are several variational principles which provide upper bounds for the Helmholtz
free-energy function.  With these instruments, it is possible to construct various approximations
to the statistical thermodynamic behavior of systems.
For any   variational formulation,  its effectiveness as a minimal principle will be enhanced considerably
if there is  a workable tool for determining lower bounds to the Helmholtz free energy function.
Bogoliubov  inequality for the free-energy functional is an inequality that gives rise to a variational
principle of statistical mechanics. It is used~\cite{bogocp12,nbog94,bb09,bb14,tyab} to obtain the exact thermodynamic limit~\cite{kuz10} solutions of model
problems in statistical  physics, in studies using the method of molecular fields, in
proving the existence of the thermodynamic limit~\cite{kuz14}, and also in order to obtain physically important
estimates for the free energies of various many-particle interacting systems.\\
A clear formulation of the variational principle of Bogoliubov and Bogoliubov  inequality for the free-energy functional
was carried out by S. V. Tyablikov~\cite{tyab}. We shall follow close to that formulation. Tyablikov~\cite{tyab}
used the theorems relating  to the minimum values of the free energy. As a result,  it was possible to
formulate a principle which then was used to deduce the molecular field equations.\\
Principle of the free energy minimum is based on the following arguments. Let us consider an arbitrary complete
system of orthonormalized functions $\{\varphi_{n} \}$, which are not the eigenfunctions of the Hamiltonian $\mathcal{H}$
of a system. Then it is possible to write down the inequality
\begin{equation}
\label{bieq1}
 F (\mathcal{H}) \leq F_{\textrm{mod}} (\mathcal{H}).
\end{equation}
Here $F (\mathcal{H})$ is the intrinsic free energy of the system
\begin{equation}
 F (\mathcal{H}) = - \theta \ln Z, \quad Z = \sum_{\nu} \exp (- E_{\nu}/\theta),
\end{equation}
$\theta =k_{B}T$, $E_{\nu}$ are eigenfunctions of the Hamiltonian $\mathcal{H}$, $F_{\textrm{mod}} (\mathcal{H})$
is the \emph{model} free energy, which gives approximately the upper limit of the 
intrinsic free energy:
\begin{equation}
 F_{\textrm{mod}} (\mathcal{H}) = - \theta \ln Z_{\textrm{mod}}, \quad Z_{\textrm{mod}} = \sum_{n} \exp (- \mathcal{H}_{nn}/\theta), 
 \quad \mathcal{H}_{nn} =  \Bigl ( \varphi^{*}_{n},   \mathcal{H} \varphi_{n} \Bigr ).
\end{equation}
The inequality (\ref{bieq1}) may be written also in the following way:
\begin{equation}
\label{bieq4}
Z   \geq  Z_{\textrm{mod}}.
\end{equation}
The relationships represented by the equality sign in Eqs. (\ref{bieq1}) and (\ref{bieq4}) applies if $\varphi_{n}$
are eigenfunctions of the Hamiltonian of the system. It should be noted that for finite values of the number of 
partial sums $Z^{(N)}$,
the quantity $F^{(N)}_{\textrm{mod}}$ does not reach its maximum for any system of functions $\varphi_{1}, \ldots,  \varphi_{N}$.
In fact, the inequality will be  satisfied really~\cite{tyab,kuz14} in the limit $N \to \infty$.\\
Using these results, it is possible to formulate a variational principle for the approximate determination of the free
energy of a system~\cite{tyab}. To proceed, let us suppose that the functions $\{\varphi_{n} \}$ depend on some
arbitrary parameter $\lambda$. It was established above that
\begin{equation}
 F (\mathcal{H}) \leq F_{\textrm{mod}} (\mathcal{H}) = - \theta \ln \sum_{n} \exp (- \mathcal{H}_{nn}(\lambda)/\theta). 
\end{equation}
It is clear that the best approximation for the upper limit of the free energy $F$ is obtained by selecting the values
of the parameter $\lambda$ in accordance with the condition for the minimum of the model free energy $F_{\textrm{mod}}$.
Indeed, let the Hamiltonian of the system, $\mathcal{H}$, be written in the form
\begin{equation}
 \mathcal{H} = \mathcal{H}_{0}(\lambda) + \Delta \mathcal{H}(\lambda) \equiv \mathcal{H}_{0}(\lambda) +  \Bigl ( \mathcal{H} - \mathcal{H}_{0}(\lambda) \Bigr ),
\end{equation}
where $\mathcal{H}_{0}(\lambda)$ is some operator depending on the parameter $\lambda$. The concrete form of the 
operator $\mathcal{H}_{0}(\lambda)$ should be selected on the basis of convenience in calculations. We shall use
notation $E_{n}^{0}$ and $\varphi_{n}$  for the eigenvalues  and the eigenfunctions of the 
operator $\mathcal{H}_{0}$.  To denote the diagonal matrix elements of the operator $\Delta \mathcal{H}$ in terms of the functions 
$\varphi_{n}$ we shall use the notation $\Delta \mathcal{H}_{nn}$.\\
For a generality, we shall assume that $\varphi_{n}$ are not the eigenfunctions of the total Hamiltonian $\mathcal{H}$.
Clearly, $E_{n}^{0}$ and $\Delta \mathcal{H}_{nn}$ are also some functions of the parameter $\lambda$. In this sense, the
system of functions $\{\varphi_{n} \}$ plays a role of a \emph{trial} system  of functions. Then we may write that
\begin{equation}
 \mathcal{H}_{nn} = E_{n}^{0} + \Delta \mathcal{H}_{nn} \equiv E_{n}^{0} + \Bigl ( \mathcal{H}_{nn} - E_{n}^{0} \Bigr ).
\end{equation}
As a consequence, the free energy will satisfy the inequality
\begin{equation}
 F (\mathcal{H}) \leq - \theta \ln \sum_{n} \exp  - \Bigl ( E_{n}^{0} + \Delta \mathcal{H}_{nn} \Bigr ) \frac{1}{\theta}.
\end{equation}
Now let us suppose that the operator $\Delta \mathcal{H}$ can be considered as a \emph{small perturbation} compared
with the operator $\mathcal{H}$. We obtain then~\cite{tyab}, to within quantities of the first order of smallness with
respect to $\Delta \mathcal{H}$
\begin{equation}\label{bieq23}
 F (\mathcal{H}) \leq   F (\mathcal{H}_{0}) + \frac{\textrm{Tr} \Bigl (\Delta \mathcal{H} \exp (- \mathcal{H}_{0}/\theta) \Bigr )}{\textrm{Tr} \Bigl ( \exp (- \mathcal{H}_{0}/\theta) \Bigr )}.
\end{equation}
Note that in this case, the best approximation to the upper limit of the free energy is obtained by selecting 
the value of the parameter $\lambda$ from the condition for the minimum of the right-hand side of Eq.(\ref{bieq23}).
The formulation of the variational principle of Eq.(\ref{bieq23}) is more restricted than the initial formulation of 
Eq.(\ref{bieq1}).\\
The variational principle in the form of Eq.(\ref{bieq23}) can be strengthened, following to the Bogoliubov suggestion~\cite{tyab},
by removing the limitation of the smallness of the operator $\Delta \mathcal{H}$. As a result we obtain
\begin{equation} 
  F (\mathcal{H}) \leq F_{\textrm{mod}} (\mathcal{H}).
\end{equation}
Here
\begin{equation} 
F_{\textrm{mod}} (\mathcal{H}) =   F (\mathcal{H}_{0})  + \frac{\textrm{Tr} \Bigl (\Delta \mathcal{H} \exp (- \mathcal{H}_{0}/\theta) \Bigr )}{\textrm{Tr} \Bigl ( \exp (- \mathcal{H}_{0}/\theta) \Bigr )},
\end{equation}
\begin{equation} 
  F (\mathcal{H}_{0}) = - \theta \ln \textrm{Tr} \exp (- \mathcal{H}_{0}/\theta).
\end{equation}
Hence one may write down also that for a system with the Hamiltonian
\begin{equation}
\mathcal{H}=\mathcal{H}_{0}+\Delta \mathcal{H} 
\end{equation}
the free energy has  a certain upper bound. Bogoliubov inequality states that:
\begin{equation}\label{bineq5}
  F \leq F_{0} +  \langle \mathcal{H}  - \mathcal{H}_{0} \rangle_{0}, 
\end{equation}
or
\begin{equation}\label{bineq6}
  F \leq F_{0}  \langle \mathcal{H} \rangle_{0} - T S_{0}, 
\end{equation}
where $S_{0}$ is the entropy and where the average is taken over the equilibrium ensemble 
of the reference system with Hamiltonian $\mathcal{H}_{0}$. 
Usually $\mathcal{H}_{0}$ contains one or more variational parameters which are chosen such as to minimize
the right hand side of Eq.(\ref{bineq5}).
In 
the special case that the reference Hamiltonian is that of a non-interacting system and can thus be written as
a sum single-particles Hamiltonians~\cite{tyab}
\begin{equation}
\mathcal{H}_{0} = \sum_{i=1}^{N}h_{i}.  
\end{equation}
Then it is possible  to improve the upper bound by minimizing the 
right hand side of the inequality (\ref{bineq5}). The minimizing reference system is then the 
trial approximation to the true system using non-correlated degrees of freedom, and is known as 
\emph{the mean field approximation}.\\
Starting with
the one-particle model Hamiltonian that can be exactly solved in the Bogoliubov
variational method, one may get a self-consistent result such as the molecular field
theory in the ferromagnet  and the Hartree-Fock approximation in many-particle problems.
Since the variational method yields a result which is always greater than
the correct answer, the mathematical meaning for improving upon the approximation
in the variational method is strictly defined by lowering the upper bound of the
free energy. But these variational methods, the molecular field theory and the
Hartree-Fock approximation, have such a feature that the correlation effects
cannot be taken into account correctly.   
In general case~\cite{tyab}   the  Hamiltonian of a system contains  
interparticle interactions. Thus Bogoliubov  variational principle can be considered as the   mathematical 
foundation of the mean field approximation in the theory of many-particle interacting systems.\\
Using  the Klein   inequality (\ref{klemm})  it is possible to write down a general form of the Bogoliubov  
inequality for the free energy functional.
The following inequality is valid for any Hermitian operators and $H_{1}$ and $H_{2}$
\begin{equation}
 N^{-1} \langle H_{1} - H_{2} \rangle_{H_{1}} \leq \Bigl ( f(H_{1}) - f(H_{2}) \Bigr ) \leq  N^{-1} \langle H_{1} - H_{2} \rangle_{H_{2}},
\end{equation}
where 
\begin{equation}
 f(H) = - \theta  N^{-1}\ln \textrm{Tr} \exp (- H/\theta).
\end{equation}
This expression has the meaning of the free  energy density for a system with Hamiltonian $H$ and the extensive
parameter $N$ may be treated as the number of particles or the volume, depending on the system.\\
Derrick~\cite{derr64} established a simple variational bound to the entropy $S(E)$ of a system with energy $E$
\begin{equation}
 S(E) \geq - k_{B} \ln \Bigl ( \textrm{Tr} U^{2} \Bigr )
\end{equation}
for all Hermitian matrices $U$ (with no negative eigenvalues) for which $\textrm{Tr} U = 1$ and $\textrm{Tr}(H U)= E$,
where $H$ is the Hamiltonian. \\ This principle has the advantage that $U^{2}$ is in general much easier to evaluate than
$U \ln U$ which appears in the conventional bound given by von Neumann
\begin{equation}
 S(E) \geq - k_{B}   \Bigl ( \textrm{Tr} U \ln U \Bigr ).
\end{equation}
There are numerous methods for proving of the Bogoliubov inequality~\cite{tyab,huber68,huber73,mazo74,deco04,hbcal,feysm72,prat96}.
A. Oguchi~\cite{ogu76} proposed an approach  for determination of an upper bound and a lower bound of the Helmholtz free energy 
in the statistical physics. He used as a basic tool the Klein lemma~\cite{carl09,rusk02,klein32}.
He obtained a new approximate expression of the free energy. This approximate value of the free energy was conjectured 
to be greater than the lower bound and less than the upper bound. An  approach which can be extended to improve 
the approximation was formulated. The upper bound and the lower bound of the approximate free energy converge to the true 
free energy as the successive approximation proceeds. 
The method was first applied to the Ising ferromagnet and then applied to the Heisenberg ferromagnet. 
In the simplest approximation the results agrees with the Bethe-Peierls approximation for the Ising model and the constant 
coupling approximation for the Heisenberg model. 
In his subsequent paper, A. Oguchi~\cite{ogu84} formulated
a new variational method for the free energy in statistical physics. According to his calculations, the value of the free energy was obtained 
by using this new variational method was lower than that of the Bogolyubov variational method. Author concluded that the new variational 
free energy satisfies the thermodynamic stability criterion.\\
However, J. Stolze~\cite{stol85} by careful examination of the papers~\cite{ogu76,ogu84}, has
shown   that the calculation in Ref.~\cite{ogu84} contains a mistake which invalidates the result. He
also pointed out several errors seriously affecting the results of an earlier paper~\cite{ogu76}.
Oguchi assumed that the Hamiltonian $\mathcal{H}_{0}$ contains a variational parameter "$a$" distributed
according to a probability density $P(a)$. 
Stolze  derived a  corrected inequality which clearly states
that the new upper bound on the free energy suggested by Oguchi~\cite{ogu76,ogu84}
cannot be better (i.e., lower) than the Peierls-Bogoliubov bound, no matter how cleverly $P(a)$
was chosen. This  shows clearly  that no advantage over the Peierls-Bogolyubov bound was obtained.\\
The standard proof was given in Callen  second edition book on thermodynamics~\cite{hbcal} for the case when
the unperturbed Hamiltonian and the perturbation commute. Another proof (for the general case), was carried out
in Feynman  book on statistical mechanics~\cite{feysm72}.
Feynman used Baker-Campbell-Hausdorff expansion~\cite{lieb03,carl09} for the exponential of a sum
of two non commuting operators. Prato and  Barraco~\cite{prat96} presented a proof of the Bogoliubov inequality that does not 
require of the Baker-Campbell-Hausdorff expansion.\\
Several variational approaches for the free energy have been proposed~\cite{silva83,ferre87} as attempts to
improve results obtained through the well established Bogoliubov principle. This
principle requires the use of a trial Hamiltonian depending on one or more variational
parameters. The only way to improve the Bogoliubov principle by itself is to choose
a more complete trial Hamiltonian, closing it to the exact one, but in almost all cases
the possibilities are soon exhausted. The usual mean field approximation may be
obtained using the above principle utilizing a sum of single spins in an effective field
(the variational parameter) as the trial Hamiltonian.\\
Lowdin~\cite{polow88} and Lowdin and Nagel~\cite{nagel89}   studied
a generalization of the Gibbs-Bogoliubov inequality $ F \leq F_{0} +  \langle \mathcal{H}  - \mathcal{H}_{0} \rangle_{0}$ 
for the free energy $F$  which leads to a variation principle 
for this quantity that may be of importance in certain computational applications to quantum systems. This approach is 
coupled with a study of the perturbation expansion of the free energy for a canonical ensemble 
with $\mathcal{H} = \mathcal{H}_{0} + \lambda V$ in the general case when $\mathcal{H}_{0}$  and $V$ do not commute.
A simple proof was given for the thermodynamic inequality 
$F - F_{0} - \langle \mathcal{H}  - \mathcal{H}_{0} \rangle_{0} < 0$  in the case when the two Hamiltonian  $\mathcal{H}_{0}$ 
and $V$ do not commute. 
The second- and high-order derivatives of the free energy with respect to the perturbation parameter $\lambda$ were 
calculated. From the second-order term was finally obtained a second-order correction to the previous variational minimum 
for the free energy.\\
A. Decoster~\cite{deco04}
established a sequence of inequalities which generalize the the Gibbs-Bogoliubov inequality in classical statistical  
mechanics and the Peierls and Bogoliubov inequalities in quantum mechanics; 
they can be presented as rearrangements of perturbation expansions, which provide exact bounds 
which are used in variational calculations.\\ 
W. Kramarczyk~\cite{kram74} argued that
the Bogoliubov variational principle may be shown to be equivalent to the minimizing of the information gained 
while replacing the exact state by an approximate one. Consequently, the quasiparticles introduced in the 
thermal Hartree-Fock approximation may be redefined information-theoretically. 
%
%
%
\section{Applications of the Bogoliubov  Variational Principle} 
%
Bogoliubov  variational principle  has been successfully applied to
a wide range of problems in the theory of many-particle systems.
The first application of Bogoliubov inequality to concrete many-particle problem was carried out in
the work by I. A. Kvasnikov~\cite{kvas56} on the application of a variational principle to the Ising model of ferromagnetism.\\
Ising model~\cite{kuz09,salin01} is defined by the following Hamiltonian  $\mathcal{H}$ (i.e. energy functional of variables; 
in this case the "spins"  $S_{i} = \pm 1$  on the $N$ sites of a regular lattice in a space of dimension $d$)
\begin{equation}
\mathcal{H} =  - \frac{1}{2} \sum_{i < j=1}^{N} J_{ij} S_{i}S_{j} - \mu_{B} H \sum_{i =1}^{N} S_{i}
 \end{equation}
Here, $J_{ij}$  play a role of  "exchange constants", $H$ is a (normalized) magnetic field, involving an interpretation 
of the model to describe magnetic ordering in solids ($\mathbf{M} = \sum_{i =1}^{N} S_{i}$ is "magnetization"; 
$\mu_{B} H \mathbf{M}$  is  the Zeeman energy, i.e. is the energy gained due to application of the field).\\ 
The main task is to calculate statistical sum $Z$
\begin{equation}
Z =  \sum_{S_{i}} \exp  - \Bigl ( \mathcal{H}/\theta  \Bigr ).  
\end{equation}
Kvasnikov~\cite{kvas56} considered the approximation of nearest neighbours, i.e. $J_{ij} = J$ for nearest neighbours
$<i,j>$.\\
According to Bogoliubov  variational principle one can write
\begin{equation}\label{dopieqn}
  F \leq F_{0} +  \langle \mathcal{H}  - \mathcal{H}_{0} \rangle_{0}. 
\end{equation}
The upper bound for the free energy $F_{\textrm{sup}}$ is given by
\begin{equation}\label{bsup}
 F_{\textrm{sup}} = - \theta \ln Z_{\textrm{inf}},
\end{equation}
where
\begin{eqnarray}
Z_{\textrm{inf}} =  Z_{0} \exp  -  ( S/\theta ),\quad  Z_{0} =  \textrm{Tr} \exp  - \Bigl ( \mathcal{H}_{0}/\theta  \Bigr ); \\
S = (Z_{0})^{-1} \textrm{Tr} \Bigl ( \Delta \mathcal{H}   \exp  -  [\mathcal{H}_{0}/\theta ] \Bigr ).
\end{eqnarray}
The parameters of partition, which were introduced into $\mathcal{H}_{0}$ and $\Delta \mathcal{H}$, and, hence, into
$Z_{0}$ and $S$, should be determined from the condition of the minimum of $ F_{\textrm{sup}}$. Thus we obtain
\begin{eqnarray}
 - \frac{\mathcal{H}_{0}}{\theta} = \mu_{B}(B -  \chi)  \sum_{i =1}^{N} S_{i},\\
- \frac{\Delta \mathcal{H}}{\theta} = \mu_{B} \chi \sum_{i =1}^{N} S_{i} + \frac{1}{2}  \sum_{i \neq j}^{N} K_{ij} S_{i}S_{j},  
\end{eqnarray}
where $B = H/\theta$, $K = J/\theta$, $\chi$ is some parameter. Then, according to relation (\ref{bsup}), one finds
\begin{eqnarray}
\Bigl ( Z_{\textrm{inf}} (\chi)  \Bigr )^{-1} = \\ \nonumber
2 \coth \mu_{B}(B -  \chi)\exp \Bigl \{ \mu_{B} \chi \tanh \mu_{B}(B -  \chi) + 
\frac{1}{2 N}  \sum_{i \neq j}^{N} K_{ij} \tanh^{2} \mu_{B}(B -  \chi) \Bigr \}.
\end{eqnarray}
Parameter $\chi$ is determined by the equation
\begin{eqnarray}\label{eq7}
 \tanh \mu_{B}(\chi - B) = \frac{N}{\sum K_{ij}} \mu_{B} \chi, \quad 
 1 - \frac{1}{N}\sum K_{ij} + \frac{N}{\sum K_{ij}} (\mu_{B} \chi)^{2} > 0.
\end{eqnarray}
When the approximation of nearest neighbours is considered in the above equation the following substitution
should be done
\begin{equation}
 \sum_{i \neq j}^{N} K_{ij} = z K N,
\end{equation}
where $z$ is the number of nearest neighbours.\\
Hence $F_{\textrm{sup}}$ is an approximate expression for the free energy and $Z_{\textrm{inf}}$ is the approximate
statistical sum of the model. It will be of instruction to compare these values with those which were calculated by other methods.
To proceed, let us consider the regions of low and high temperatures. In the first case we will have that $\theta \ll z J$.
The low-temperature approximation is expressed as  a  series expansion in terms of the small parameter $\exp (- K)$. The iterative
solution of the Eq.(\ref{eq7}) will have the form
\begin{equation}
 \mu_{B} \chi = - z K \Bigl ( 1 - 2 \exp 2(- K z - \mu_{B} B) - 8 z K \exp 4(- K z - \mu_{B} B) + \ldots  \Bigr ).
\end{equation}
It is sufficient to confine oneself to the values of the order $\exp (- 2 K z)$. The result is
\begin{equation}
  \Bigl ( Z_{\textrm{inf}} \Bigr )^{-1} =   \exp (  K z/2 + \mu_{B} B)   \Bigl ( 1 + \exp 2(- K z -  B) + \ldots \Bigr ).
\end{equation}
This result  is in accordance with the other low-temperature expansions~\cite{tyab,salin01}
\begin{eqnarray}
 Z = \exp 2(  K z/2 + \mu_{B} B)N \Bigl ( 1 + N \exp 2(- K z - \mu_{B} B) + \frac{N z}{2} \exp 4[- K (z - 1) - \mu_{B} B]\nonumber \\
 + \{ \frac{N(N - 1)}{2}  - \frac{N z}{2}\} \exp 4(- K z - \mu_{B} B) + \ldots  \Bigr ).
\end{eqnarray}
In the case of the high temperature, when $\theta \geq z J$, the approximate solution of  the Eq.(\ref{eq7}) will have the form
\begin{equation}
  \mu_{B} \chi \simeq - z K \frac{\tanh \mu_{B} B}{1 - [ z K/\cosh^{2} \mu_{B} B]}.
\end{equation}
Then after some transformations one can arrive to the expression (up to the terms $K^{3}$):
\begin{eqnarray}
 Z_{\textrm{inf}} \simeq [2 \cosh \mu_{B} B]^{N} \Bigl ( 1 + \frac{1}{2} K z \tanh^{2} \mu_{B} B + \\ \nonumber
\frac{1}{8} K^{2} z N [4 z \tanh^{2} \mu_{B} B + (N z + 4 z)\tanh^{4} \mu_{B} B ] \Bigr ).
\end{eqnarray}
This expression is also in accordance with the known high-temperature expansions~\cite{tyab,salin01} for $N \gg z$.\\
Let us consider now the expression for magnetization~\cite{grif66} (the averaged magnetic moment)
\begin{equation}
 M = \frac{1}{N} \frac{\partial \ln Z_{\textrm{inf}}}{\partial B}.
\end{equation}
Using Eq.(\ref{eq7}), we obtain
\begin{equation}
 \frac{m}{\mu_{B} p} = \tanh \mu_{B} \Bigl ( \frac{H}{\theta}  + n \frac{m}{\theta} \Bigr ),
\end{equation}
where $p$ is the number of lattice sites per unit volume, $m = M p$ is the magnetization per unit volume.
This result coincides with the result of the phenomenological theory~\cite{tyab}. The corresponding basic values
of the P. Weiss theory, the Curie point $\theta_{0}$  and Weiss parameter $w$  have the form
\begin{equation}
 \theta_{0} = \frac{1}{N} \sum_{i \neq j}^{N} J_{ij}; 
 \quad w = \frac{N^{-1} \sum J_{ij}}{\mu_{B}^{2} p} = \frac{\theta_{0}}{\mu_{B}^{2} p}.
\end{equation}
Hence, with the help of the Bogoliubov variational scheme it was possible to calculate the reasonable
approximate expression for the statistical sum of the Ising model and describe the macroscopic properties
of ferromagnetic systems in the wide interval of temperatures. 
It is thus seen that one may derive directly a consistent mean-field-type theory from a variational principle.\\
Clearly Bogoliubov variational principle had a deep impact on the field of statistical mechanics of classical and
quantum many-particle systems by making possible the analysis of complex statistical systems.
Many interesting developments can be viewed from the point of a central
theme, namely the Bogoliubov inequality, in particular in quantum theory of magnetism~\cite{tyab,grif66,radc68,rudo70,ogu70,solo91}
and interacting many-body systems~\cite{tolm60,rich69,lieb81,bach14,ogu87,wern91,breto92,vlach93,bach14a}.\\
Radcliffe~\cite{radc68} carried out
a systematic investigation  of the approximate free energies and Curie temperatures that can be
obtained by using trial density matrices (which describe various possible decompositions of the ferromagnet
into clusters) in a variational calculation of the free energy. Single-spin clusters lead to the molecular field
model (as is well known) and two-spin clusters yield the Oguchi pair model~\cite{smart}. The relation of the
\emph{constant-coupling}  method to these approximations was clarified. A rigorous calculations using three-spin
clusters were carried out. \\ 
Rudoi~\cite{rudo70} investigated
the link between Bogoliubov  statistical variational principle for free energy, the method
of partial diagram summation of the perturbation theory, and the Luttinger-Ward theorem.
On the basis of Matsubara  Green  function method he solved the nonlinear integral
Dyson equation  by approximating the effective potential.  As a result, a new implicit equation
of magnetic state was obtained for the Ising model.\\
Soldatov~\cite{soldat} generalized the Peierls-Bogoliubov inequality.  A set of inequalities was derived instead,
so that every subsequent inequality in this set approximates the quantity in question with better
precision than the preceding one. These inequalities lead to a sequence of improving upper bounds to
the free energy of a quantum system if this system allows representation in terms of coherent states.
It follows from the results obtained that nearly any upper bound to the ground
state energy obtained by the conventional variational principle can be improved by means of the proposed method.\\
F. Abubrig~\cite{abub13}   studied
the mixed spin-3/2 and spin-2 Ising ferrimagnetic system with different single-ion anisotropies in the absence of 
an external magnetic field  within the mean-field theory based on Bogoliubov inequality for the Gibbs free energy. 
Second-order critical lines were obtained in the temperature-anisotropy plane. Tricritical line separating second-order and first-order 
lines was found. Finally, the existence and dependence of a compensation points on single-ion anisotropies was also investigated for the 
system. It was shown that this mixed-spin model exhibits one, two or three compensation temperature depending on the values 
of the anisotropies. 
%
%
%
%
%
\section{The Variational Schemes and Bounds on Free Energy} 
%
%
%
During last few decades, numerous variational schemes have become an increasingly popular workable tool in quantum-mechanical 
many-particle theory~\cite{tyab,huber68,huber73,girar90}.
Bounds of free energy and canonical ensemble averages were of considerable interest as well.  For many
complex systems, such as Ising and Heisenberg ferromagnets  or
composite materials,  methods of obtaining bounds are the practical useful tools which are both tractable and
informative. A few illustrative topics will be of instruction to discuss in this context.\\
MacDonald and Richardson~\cite{macdon54} used
the density matrix of von Neumann   to formulate an exact variational principle for quantum statistics which 
embodies the principle of maximization of entropy. In terms of the formalism of second quantization, authors  wrote this 
variational principle for fermions or bosons and   then derived from it an approximate variational procedure which 
yields the particle states of a system of interacting bosons or fermions as well as the distribution of particles in 
these states. These equations, in authors's opinion,  yield the generalization of the 
Hartree-Fock equations for nonzero temperature and the corresponding extension to bosons.\\
W. Schattke~\cite{schat68} found an upper bound for the free energy for superconducting system in magnetic field.
Starting from the BCS  theory,  the free energy was obtained by a combination
of a variational method and perturbation theory. The variational equations obtained were non-local.
The parameters of the perturbation calculation were the vector potential and the spatial variations
of the order parameter, which have to be small. Boundary conditions were set for the case of diffuse
reflection and pair-breaking at the surface. As an example, the superconducting plate was discussed.\\
Krinsky et. al. used~\cite{blum74} the variational principle to derive a new approximation to a ferromagnet in
a magnetic field, below its critical temperature. They considered~\cite{blum74}
a ferromagnet in an external magnetic field with  $T \leq  T_{C} $. Using a variational approximation based on the 
zero-field solution, the authors  obtained an upper bound on the free energy, an approximate equation of state, 
and a lower bound on the magnetization, all having the correct critical indices. 
Explicit numerical calculations have been carried out for the two-dimensional Ising model, and it was found that 
the results obtained provide a good approximation to the results of series expansions throughout the 
region $T \leq  T_{C} $.\\ 
The Gibbs-Bogoliubov inequality~\cite{ishi68} was used~\cite{stel70} to develop a first-order perturbation theory that provides an 
upper bound on the Helmholtz free energy per unit volume of a classical statistical mechanical system in terms of the 
free energy and pair distribution function. Charged systems as well as a system of Lennard-Jones particles were
discussed and detailed numerical estimates of the bounds were presented.\\
S. Okubo and A. Isihara~\cite{okubo72a}
derived important general inequalities    for the derivatives of the partition function of a quantum system with respect 
to the parameters included in the Hamiltonian. Applications of the inequalities were used to discuss relations for critical 
initial exponents, kinetic energy, susceptibility, electrical conductivity and so on. Existence of an inconsistency 
analogous to the Schwinger-term difficulty in the quantum field theory was pointed out.\\
In their second paper~\cite{okub72},  S. Okubo and A. Isihara analyzed  from a general point of view an 
inequality for convex functions in quantum-statistical mechanics. From an inequality for a convex function of two hermitian 
operators, the Peierls and Gibbs operators, coarse graining and other important 
inequalities were derived in a unified way. Various different forms of the basic inequality were given. They are found 
useful in discussing the entropy and other physical problems. Special accounts were given of functions such 
as  $\exp (x)$  and  $x \log x$.\\
A variational method for many-body systems using a separation into a difference of Hamiltonians was presented by 
M. Hader and F. G. Mertens~\cite{hade88}.
A particular ansatz for the wave function  was considered which leads to an upper bound for the exact ground-state energy. 
This allowed a variation with respect to a separation parameter. The method was tested for a one-dimensional lattice 
with Morse interactions where the Toda subsystems can be solved by the Bethe ansatz. In two limiting cases  results obtained 
were exact, otherwise they were in  agreement with the quantum transfer integral method.\\
Yeh~\cite{yeh69,yeh70,yeh71} proposed a derivation of a lower bound on the free energy; in addition he  analyzed
the bounds of the average value of a function~\cite{yeh71}. He also  established~\cite{yeh69}  
a weaker form of Griffiths  theorem  for the ferromagnetic Heisenberg model. It was described as follows~\cite{yeh70}: 
Free energy in the canonical ensemble was taken as
\begin{equation}
 F = - \beta^{-1}  \ln \sum_{n} \langle n|  \exp ( - H  \beta)|n \rangle, 
\end{equation}
where ${|n \rangle}$ is any complete set of orthonormal states. Bounds of
$F$ can be obtained from bounds of $\langle n|  \exp ( - H  \beta)|n \rangle$. As we seen, a very simple upper bound 
of $F$ was given by Peierls~\cite{peier38}; one way  to prove his theorem is by showing that
\begin{equation}
\langle \psi|  \exp ( - H  \beta)|\psi \rangle \geq   \exp ( -   \beta \langle \psi|  H  |\psi \rangle).
\end{equation}
Yeh~\cite{yeh70} derived a rather simple lower bound of $F$ by similar method. He considered a 
a Hamiltonian with a ground-state energy $E_{0} = 0$. He considered a real function $f(E) = \exp ( - E  \beta)$, $\beta > 0$.
It was  shown that for any normalized state $|\psi \rangle$
a weaker  but simpler upper bound for $f$  may be written as
\begin{equation}\label{yeh3}
\exp \Bigl ( -   \beta \langle \psi|  H  |\psi \rangle p \Bigr ) \geq \langle \psi|  \exp \Bigl ( - H  \beta \Bigr )|\psi \rangle \geq 
\exp \Bigl ( -   \beta \langle \psi|  H  |\psi \rangle \Bigr ),
\end{equation}
where
\begin{equation}
 p =    \exp \Bigl (  \frac{( -   \beta \langle \psi|  H^{2}  |\psi \rangle)}{\langle \psi|  H  |\psi \rangle} \Bigr )
\end{equation}
Identifying $\beta = (k_{B}T)^{-1}$ and $H$ as Hamiltonian,  a lower
bound of free energy was obtained  from Eq.(\ref{yeh3}) as
\begin{equation}
 F \geq - \beta^{-1}  \ln \sum_{\psi}   \exp \Bigl (  -   \beta \langle \psi| H  |\psi \rangle  p \Bigr ),
\end{equation}
where ${|\psi \rangle}$ is any complete orthonormal set of states. This is a general formula for a lower bound on the free energy.\\
Upper and lower bounds of the canonical ensemble
average of any operator $A$ can be written down in
terms of  $ \langle \varphi_{n}| H  |\varphi_{n} \rangle$, where the $\varphi_{n}$ are eigenstates
of $A$. Furthermore, bounds of thermodynamic derivatives
can be obtained by noting that the bounds of
\begin{equation}
 \frac{\partial^{i} \bar{f} }{\partial \beta^{i}}
\end{equation}
can be also derived~\cite{yeh70} in similar manner. Here 
\begin{equation}
 \bar{f} = \langle \psi|  \exp ( - H  \beta)|\psi \rangle = \sum_{n} \rho_{n}   \exp ( - E_{n}  \beta); \quad \sum_{n} \rho_{n} = 1.
\end{equation}
From Eq.(\ref{yeh3}), it is clear that all the bounds 
are more accurate at higher temperatures. These bounds have been useful
in determining the properties of Heisenberg ferromagnets~\cite{yeh69}.\\
K. Symanzik~\cite{syma65} proved, refined, and generalized 
a lower bound given by Feynman for the quantum mechanical free energy of an oscillator.  
The method, application of a classical inequality to path integrals, also gives upper bounds for 
one-temperature Green functions.\\
M. Heise and R. J. Jelitto~\cite{heis76} formulated  the asymptotically exact variational approach to the strong coupling 
Hubbard model.
They used a generalization of Bogoliubov  variational principle, in order
to develop a molecular field theory of the Hubbard model, which becomes asymptotically
exact in the strong coupling limit. 
In other words, in their paper authors have started from a generalized
form of Bogoliubov  variational theorem in order to set up a theory of the Hubbard model, which
yields nontrivial results in the strong coupling regime
and becomes asymptotically exact in the strong coupling limit. For this purpose the Hamiltonian was rotated by a unitary
two-particle transformation, before the variational principle was applied. However, the real form of the generalized
mean fields for the the Hubbard model in the strong coupling regime was not determined in complete form. This task was
fulfilled by A. L. Kuzemsky in a series of papers~\cite{kuz09,kuz78,kuznc94,kuzrnc02}.\\
K. Zeile~\cite{zei78} proposed
a generalization of Feynman variational principle for real path integrals in a systematic way.
He obtained an asymptotic series of lower bounds for the partition function. 
Author claimed that the method was tested on the anharmonic oscillator and showed excellent agreement with 
exact results. 
However, P. Dorre~\cite{dorr79} et al. using the equivalence between Feynman  and Bogoliubov  variational principle, 
discussed~\cite{dorr79} in the formalism of Hamiltonian quantum mechanics an improved upper bound for the free energy  
 which has been given  by Zeile~\cite{zei78} using path integral methods. It was shown 
that Zeile's variational principle does not guarantee a thermodynamically consistent description.\\
U. Brandt and J. Stolze formulated~\cite{brand81} 
a new hierarchy of upper and lower bounds on expectation values.
Upper and lower bounds were constructed for expectation values of functions of a real
random variable with derivatives up to order $(N + 1)$ which are alternately negative and
positive over the whole range of interest. The bounds were given by quadrature formulas
with weights and abscissas determined by the first $(N + 1)$ moments of the underlying
probability distribution. Application to a simple disordered phonon system yielded sharp
bounds on the specific heat.\\
K. Vlachos~\cite{vlach93} proposed
a variational method that uses the frequency and the energy shift as variational parameters. 
The quantum-mechanical partition function was approximated by a formally simple expression, for a generalized anharmonic 
oscillator in one and many dimensions. The numerical calculations for a single quartic and two coupled quartic oscillators 
have led to nearly exact values for the free energy, the ground state, and the difference between the ground state and 
the first excited state.\\
C. Predescu~\cite{prede02} presented
a generalization of the Gibbs-Bogoliubov-Feynman inequality for spinless particles  and then illustrated it for the simple 
model of a symmetric double-well quartic potential. The method gives a pointwise lower bound for the finite-temperature 
density matrix and it can be systematically improved by the Trotter composition rule. It was also shown to produce ground 
state energies better than the ones given by the Rayleigh-Ritz principle as applied to the ground state eigenfunctions of 
the reference potentials. Based on this observation, 
it was conjectured that the \emph{local variational principle}  may performs better than the equivalent methods based on the 
centroid path idea and on the Gibbs-Bogoliubov-Feynman variational principle, especially in the range of low temperatures.
However clear evidence for such a statement was not given.\\
All these points of view acquire significance of the variational principles as a general method of solution for 
better insight into the complicated behavior of the many-particle systems.
%
%
%
%
\section{The  Hartree-Fock-Bogoliubov Mean Fields} 
%
%
%
The microscopic theory of superconductivity was created simultaneously by   Bardeen,
Cooper, and Schrieffer~\cite{bcs57,bari60}  and Bogoliubov~\cite{bozubt57,nnb58,btsh58,bogufn59,nnbo60,bozubt60,nnb71}. 
An important contribution to the theory of superconductivity were the works of Fr\"{o}hlich~\cite{froh61}, who put
forward the idea of the importance of the electron-phonon interaction for the phenomenon of superconductivity,
and the theory of Schafroth, Butler, and Blatt~\cite{sbb60}, who conjectured that superconductivity is due to Bose-Einstein 
condensation of correlated electron pairs.
In their paper, Bardeen, Cooper, and Sehrieffer  determined the ground state 
energy, and the spectrum of elementary excitations of their model~\cite{bcs57,bari60}.
The BCS theory was constructed on the basis of a model
Hamiltonian that takes into account only the interaction of electrons with opposite momenta and spins,
whereas Bogoliubov  theory  was based on the Fr\"{o}hlich Hamiltonian~\cite{froh61} and used the method of compensation
of dangerous diagrams~\cite{btsh58}. 
N. N. Bogoliubov, V. V. Tolmachev and D. V. Shirkov~\cite{btsh58} have generalized to Fermi systems
the Bogoliubov method of canonical transformations proposed earlier in connection 
with a microscopic theory of superfluidity for Bose systems~\cite{nnb47}.  
This approach has  formed the basis of a new method for investigating the problem of superconductivity.
Starting from Fr\"{o}hlich  Hamiltonian, the energy of the superconducting ground state and the one-fermion and 
collective excitations corresponding to this state were obtained. It turns out that the final formulae for the ground 
state and one-fermion excitations  obtained independently by Bardeen, Cooper and Schrieffer~\cite{bcs57} were 
correct in the first approximation. The physical picture appears to be closer to the one proposed by 
Schafroth, Butler and Blatt. The effect on superconductivity of the Coulomb interaction between the electrons was 
analyzed in detail. A criterion for the superfluidity of a Fermi system with a four-line vertex Hamiltonian was established.\\
Roughly speaking, to explain  simply the theory of superconductivity it is possible to say that the Fermi sea is unstable against the formation of
a bound Cooper pair when the net interaction is attractive;  it is reasonable
to expect  that the pairs will be condense until an equilibrium point will be reached.
The corresponding  antisymmetric wave functions for many electrons was constructed in BCS model~\cite{lieb61,haag62}.
They noted also that their solution may be considered as an exact in the  thermodynamic limit. \\ 
The most clear and rigorous arguments in favor of the statement that  the BCS model is an exactly solvable model 
of statistical physics were advanced in the papers of Bogoliubov, Zubarev, and Tserkovnikov~\cite{bozubt57,bozubt60,nnb71}. 
They showed that the free energy and the correlation functions of the BCS model and a model with a certain approximating
quadratic Hamiltonian are indeed identical in the thermodynamic limit.
In his theory~\cite{bozubt57,nnb58,btsh58,bogufn59,nnbo60,bozubt60,nnb71}, Bogoliubov gave a rigorous proof that 
at vanishing temperature the correlation functions and mean values of the energy of the BCS model and the 
Bogolyubov-Zubarev-Tserkovnikov  model are equal in the thermodynamic limit.
Moreover, Bogoliubov  constructed a complete theory of superconductivity on the
basis of a model of interacting electrons and phonons~\cite{bozubt57,nnb58,btsh58,bogufn59,nnbo60,bozubt60,nnb71}. 
Generalizing his method of canonical
transformations~\cite{petqsm95,wax94,dere07} to Fermi systems and advancing the principle of 
compensation of dangerous graphs~\cite{btsh58}, he determined the ground state consisting of paired electrons with 
opposite moments and spins, its energy, and the energy of elementary excitations. It was shown also that the phenomenon
of superconductivity consists in the pairing of electrons and a phase transition
from a normal state with free electrons to a superconducting state with pair condensate.\\
The pairing Hamiltonian has the form
\begin{equation}
\mathcal{H} - \mu \mathcal{N} =  \sum_{k\sigma} E(k) a^{\dagger}_{k\sigma}a_{k\sigma} +
\sum_{k p} V(k,p)  a^{\dagger}_{k \uparrow} a^{\dagger}_{- k \downarrow}  a_{- p \downarrow} a_{p \uparrow},
\end{equation}
where $\mu$ is the chemical potential and $\mathcal{N}$ is the number of particles.\\
The essential step which was made by Bogoliubov was connected with introducing the anomalous averages or the generalized mean fields
$F_{p} = \langle a_{- p \downarrow} a_{p \uparrow} \rangle$. It is reasonably to suppose that 
because of the large number of particles involved, the fluctuations
of $a_{- p \downarrow} a_{p \uparrow}$ about these expectations values $F_{p}$ must be small.
Hence, it is thus possible to express such products of operators in the form
\begin{equation}
 a_{- p \downarrow} a_{p \uparrow} = F_{p} + ( a_{- p \downarrow} a_{p \uparrow} -  F_{p}).
\end{equation}
It is reasonably to suppose that one may neglects by the quantities which are bilinear in the presumably small
fluctuation term in brackets. This way leads to the Bogoliubov model Hamiltonian of the form
\begin{eqnarray}
\mathcal{H}_{\textrm{mod}} - \mu \mathcal{N} = \nonumber \\ \sum_{k\sigma} E(k) a^{\dagger}_{k\sigma}a_{k\sigma}
+ \sum_{k p} V(k,p)  \Bigl ( a^{\dagger}_{k \uparrow} a^{\dagger}_{- k \downarrow} F_{p} + 
F^{*}_{k} a_{- p \downarrow} a_{p \uparrow} - F^{*}_{k}F_{p} \Bigr ).
\end{eqnarray}
Here the $F_{k}$ should be determined self-consistently~\cite{bozubt57,nnb58,btsh58,bogufn59,nnbo60,bozubt60,nnb71}. \\
Thus Bogoliubov created a rigorous theory of superfluidity~\cite{nnb47} and superconductivity~\cite{nnb71} 
within the unified scheme~\cite{petri96} of the non-zero anomalous averages or the generalized mean fields,
and showed that at the physical basis of these two fundamental phenomena of nature lies the 
process  of condensation of Bose particles~\cite{stoof97} and, respectively, pairs of fermions.\\
Indeed, N. N. Bogoliubov,  D. N. Zubarev and Yu. A. Tserkovnikov~\cite{bozubt57,bozubt60},
have   shown on the basis of the model Hamiltonian of BCS-Bogoliubov,
that the thermodynamic functions of a superconducting system, which were
obtained by a variation method in BCS, are asymptotically exact for $V \to \infty, N/V = \textrm{const}$ ($V$ is the
volume of the system, and $N$ the number of particles). This conclusion was based on the fact that
each term of the perturbation-theory series, by means of which the correction to that solution was
calculated, is asymptotically small for $V \to \infty$.
In addition, it was shown that it is possible to satisfy with asymptotic exactness the entire chain of equations
for Green  functions constructed on the basis of the model Hamiltonian of  BCS-Bogoliubov.
Thus the asymptotic exactness of the known solution for the superconducting
state was proved without the use of perturbation theory. It was shown also that the trivial
solution that corresponds to the normal state should be rejected at temperatures below the
critical temperature. In other words, starting with the  \emph{reduced Hamiltonian of superconductivity theory}, 
Bogoliubov, Zubarev, and Tserkovnikov~\cite{bozubt57,bozubt60} proved the possibility of exact calculation of the 
free energy per unit volume.  \\ Somewhat later, on the basis of the BCS theory, a similar investigation 
was made by other authors~\cite{bbmuh62,chen65,kem65,fung89}. B. Muhlschlegel~\cite{bbmuh62} studied an asymptotic expansion of the BCS partition function by 
means of the functional method.
The canonical operator $\exp [- \beta(H - \mu N)]$ associated with the BCS model 
Hamiltonian of superconductivity was represented as a functional integral by the use of Feynman's ordering parameter. 
General properties of the partition function in this representation were investigated. Taking the inverse volume of the 
system as an expansion parameter, it was possible to calculate the thermodynamic potential including terms independent 
of the volume. Muhlschlegel's theory yielded an additional 
evidence that the BCS variational value is asymptotically exact. The behavior of the canonical operator for 
large volume was described and related to the state of free quasiparticles. A study of the terms of the thermodynamic 
potential which were of smaller order in the volume in the low-temperature limit, showed that the ground state 
energy is \emph{nondegenerate} and belongs to a number eigenstate.\\
W. Thirring and A. Wehrl~\cite{wthir67}
investigated in which sense the Bogoliubov-Haag treatment of
the BCS-Bogoliubov  model gives the correct solution in the limit of infinite volume. They found
that in a certain subspace of the infinite tensor product space the field operators
show the correct time behaviour in the sense of strong convergence. Thus a solution of
the superconducting type with a gap in the spectrum of elementary
excitations  really can exists   for the model Hamiltonian of BCS-Bogoliubov.\\
In general, the problem of explaining the phenomenon of superconductivity required the solution of
the very difficult mathematical problems associated with the foundation of applied approximations~\cite{nbog94,petqsm95}.
In connection with this, Bogoliubov investigated~\cite{bozubt57,nnb58,btsh58,bogufn59,nnbo60,bozubt60,nnb71} the reduced 
Hamiltonian, in which the interaction of single electrons is studied, and carried out for it a complete mathematical
investigation for zero temperature. In this connection he laid the bases of a new powerful
method of the \emph{approximating Hamiltonian}, which allows linearization of nonlinear quantum
equations of motion, and reduction of all nonlinearity to \emph{self-consistent equations} for the
ordinary functions into which the defined operator expressions translate. This method was
extended later to nonzero temperatures and a wide class of systems, and became one
of most powerful methods of solving nonlinear equations for quantum fields~\cite{nbog94,petqsm95}.\\
D. Ya. Petrina contributed much to the further clarification of many complicated aspects of the BCS-Bogoliubov theory.
He performed a close and subtle analysis~\cite{petqsm95,petri96,petr70,petr72,petri04,petri08} of
the BCS-Bogoliubov  model and various related mathematical problems.\\
In his paper~\cite{petr70} "Hamiltonians of quantum statistics and the model Hamiltonian of the theory of superconductivity"
an investigation was made of the general Hamiltonian of quantum statistics and the model
Hamiltonian of the theory of superconductivity in an infinite volume. The Hamiltonians
were given a rigorous mathematical definition as operators in a Hilbert space of sequences
of translation-invariant functions. It was established that the general Hamiltonian is \emph{not symmetric} but possesses
a real spectrum; the model Hamiltonian is \emph{symmetric} and its spectrum has a gap between
the energy of the ground state and the excited states.\\
In the following paper~\cite{petr72},  the model Hamiltonian of the theory of superconductivity was investigated 
for an infinite volume and a complete study was made of its spectrum. The grand partition function was determined and
the equation of state was found. In addition,   the existence of a phase transition from the normal to the
superconducting state was proved. It was shown that in the limit $V \to \infty$ the chain of equations for the
Green  functions of the model Hamiltonian has two solutions, namely  the free Green  function 
and the Green  function  of the approximating Hamiltonian.\\
In his paper~\cite{petri04} D.Petrina has shown that the
Bogolyubov  result  that the average energies (per unit volume) of the ground states for the BCS-Bogoliubov Hamiltonian 
and the approximating Hamiltonian asymptotically coincide in the thermodynamic limit is
also valid for all excited states. He also established that, in the thermodynamic limit, the BCS-Bogoliubov Hamiltonian 
and the approximating Hamiltonian asymptotically coincide as quadratic forms.\\
D.Petrina~\cite{petri08}
considered  also the BCS Hamiltonian with sources, as it was proposed by Bogoliubov and Bogoliubov, Jr. It was
proved that the eigenvectors and eigenvalues of the BCS-Bogoliubov Hamiltonian with sources can be exactly determined
in the thermodynamic limit. Earlier, Bogoliubov proved that the energies per volume of the BCS-Bogoliubov
Hamiltonian with sources and the approximating Hamiltonian coincide in the thermodynamic limit.
These results clarified substantially the microscopic theory of superconductivity and provided a deeper mathematical foundation
to it.\\
Raggio and Werner~\cite{wern91} have shown the existence of the limiting free-energy density of inhomogeneous 
(site-dependent coupling) mean-field models in the thermodynamic limit~\cite{kuz14}, and derived a variational formula for this 
quantity. The formula requires the minimization of an energy term 
plus an entropy term as a functional depending on a function with values in the one-particle state space. 
The minimizing functions describe the pure phases of the system, and all cluster points of the sequence of finite 
volume equilibrium states have unique integral decomposition into pure phases. Some applications were considered; 
they include the full BCS-model, and random mean-field models.\\
A detailed and careful mathematical analysis of certain aspects of the BCS-Bogoliubov theory was carried out by 
S. Watanabe~\cite{wat08,watana08,watbe08,wata08,wata09,wat11,wat13,watab13,wata14}, mainly 
in the context of the solutions to the BCS-Bogoliubov gap equation for superconductivity.\\
BCS-Bogoliubov theory correctly yields an energy gap~\cite{vanse85,yang91}. The determination of this important energy gap is by solving a 
nonlinear singular integral equation.
An investigation of the solutions to the BCS-Bogoliubov gap equation for superconductivity was carried out 
by S. Watanabe~\cite{wat08,watana08,watbe08,wata08,wata09,wat11,wat13,watab13,wata14}. In his works the BCS-Bogoliubov equations 
were studied in full generality.
Watanabe investigated the gap equation in the BCS-Bogoliubov theory of superconductivity, where the gap function is a function of the 
temperature $T$ only. It was shown that the squared gap function is of class $C^2$ on the closed interval $[\,0,\,T_{C}\,]$. 
Here, $T_C$ stands for the transition temperature. Furthermore, it was shown that the gap function is monotonically 
decreasing on $[0,\,T_{C}]$ and  
the behavior of the gap function at $T = T_{C}$ was obtained  and some more properties of the gap function were pointed out.\\ 
On the basis of his study Watanabe then gave, by examining the thermodynamical potential, 
a mathematical proof that the transition to a superconducting state is a second-order phase transition. 
Furthermore, he obtained a new and more precise form of the gap in the specific heat at constant volume from 
a mathematical point of view. 
It was shown also that the solution to the BCS-Bogoliubov gap equation for superconductivity is continuous 
with respect to both the temperature and the energy under the restriction that the temperature is very small. 
Without this restriction,  the solution is continuous with respect to both the temperature and the energy, 
and, moreover, the solution is Lipschitz continuous and monotonically decreasing with respect to the temperature.\\ 
D. M. van der Walt, R. M. Quick  and M. de Llano~\cite{walt93,walt94}
have obtained  analytic expressions for the BCS-Bogoliubov gap of a many-electron system within the BCS model interaction 
in one, two, and three dimensions in the weak coupling limit, but for \emph{arbitrary} interaction 
width $\nu = \hbar D/E_{F}$, $0 < \nu < \infty$. 
Here $\hbar D$ is the maximum energy of a force-mediating boson, and $E_{F}$ is the Fermi energy (which is fixed by the electronic density). 
The results obtained addressed both phononic  $(\nu \ll 1)$  as well as nonphononic (e.g., exciton, magnon, plasmon, etc.) pairing 
mechanisms where the mediating boson energies are \emph{not} small compared with $E_{F}$, provided weak electron-boson coupling 
prevails. The essential singularity in coupling, sometimes erroneously attributed to the two-dimensional character of the 
BCS model interaction with $(\nu \ll 1)$,  was shown to appear in one, two, and three dimensions \emph{before} the limit $\nu \to 0$ is taken.\\
B. McLeod and Yisong Yang~\cite{leod00} studied the uniqueness and approximation of a positive solution of the 
BCS-Bogolyubov gap equation at finite temperatures. When the kernel was 
positive representing a phonon-dominant phase in a superconductor, the existence and uniqueness of a gap solution 
was established in a class which contains solutions obtainable from bounded domain approximations. The critical 
temperatures that characterize superconducting-normal phase transitions realized by bounded domain approximations and 
full space solutions were also investigated.  
It was shown under some sufficient conditions that these temperatures are identical. In this case the uniqueness of a full 
space solution follows directly. Authors~\cite{leod00}   also presented some examples for the nonuniqueness of solutions. 
The case of a kernel function with varying signs was also considered. It was shown that, at low temperatures, there exist 
\emph{nonzero gap solutions} indicating a superconducting phase, while at high temperatures, the only solution is the 
zero solution, representing the dominance of the normal phase, which establishes again the existence of a transition temperature.\\
In a series of papers~\cite{combe11,zhu12,combe13}, M. Combescot, W.V. Pogosov and O. Betbeder-Matibet studied
various aspects of the BCS ansatz for superconductivity~\cite{bcs57} in the light of the Bogoliubov approach.\\ 
In  paper~\cite{combe11}
they extended the one-pair Cooper configuration towards BCS-Bogoliubov model of superconductivity
by adding one-by-one electron pairs to an energy layer, where a small attraction acts. To do
it, they solved Richardson's equations analytically in the dilute limit of pairs on the one-Cooper pair scale.
It was found, through only keeping the first order term in this expansion, that the $N$ correlated pair energy
reads as the energy of $N$ isolated pairs within a $N (N - 1)$  correction induced by the \emph{Pauli exclusion principle}
which tends to decrease the average pair binding energy when the pair number increases. Quite
remarkably, extension of this first-order result to the dense regime gives the BCS-Bogoliubov condensation energy
exactly. These facts may lead ones to a different interpretation of the BCS-Bogoliubov condensation energy with a pair
number equal to the number of pairs feeling the potential and an average pair binding energy reduced
by Pauli blocking to half the single Cooper pair energy - instead of the more standard but far larger superconducting.\\
In the next paper~\cite{zhu12}
the usual formulation of the BCS-Bogoliubov ansatz for superconductivity in the grand canonical ensemble makes the
handling of the Pauli exclusion principle between paired electrons straightforward. It however masks that
many-body effects between Cooper pairs interacting through the reduced BCS-Bogoliubov potential are entirely controlled
by this exclusion. To show it up, one has to work in the canonical ensemble. The proper handling of
Pauli blocking between a fixed number of composite bosons is however known to be quite difficult. To do
it, author have developed a commutator formalism for Cooper pair condensate, along the line they used for excitons.
Authors~\cite{zhu12} then rederived, within the $N$-pair subspace, a few results of BCS-Bogoliubov 
theory of superconductivity obtained in the grand canonical ensemble, to evidence their Pauli blocking origin. They ended by reconsidering what should
be called  \emph{Cooper pair wave function}  and concluded differently from usual understanding.\\
In their third paper M. Combescot, W.V. Pogosov and O. Betbeder-Matibet~\cite{combe13} showed that
the Bogoliubov approach to superconductivity provides a strong mathematical support to the wave function ansatz proposed 
by Bardeen, Cooper and Schrieffer~\cite{bcs57}. However there are some subtle differences in the both the approaches. Indeed, the BCS ansatz - with all pairs condensed into the same state - corresponds to the ground state of 
the Bogoliubov Hamiltonian. From the other hand, this Hamiltonian only is part of the BCS Hamiltonian. As a result, the BCS ansatz 
definitely differs from the BCS Hamiltonian ground state. This can be directly shown either through a perturbative approach 
starting from the Bogoliubov Hamiltonian, or better by analytically solving the BCS Schrodinger equation along 
Richardson-Gaudin exact procedure. Still, the BCS ansatz leads not only to the correct extensive part of the ground state 
energy for an arbitrary number of pairs in the energy layer where the potential 
acts - as recently obtained by solving Richardson-Gaudin equations analytically - but also to a few other physical 
quantities such as the electron distribution, as it was shown by authors. The paper~\cite{combe13} also considered arbitrary 
filling of the potential layer and evidences the existence of a super dilute and a super dense regime of pairs, 
with a gap different from the usual gap. These regimes constitute the lower and upper 
limits of density-induced BEC-BCS cross-over in Cooper pair systems. It should be noted, however, that this theory needs
an additional careful examination.\\
In 1958 N. N. Bogoliubov~\cite{bogvp58} proposed a new variational principle in the many-particle problem.
This variational principle  is the generalization of the Hartree-Fock variational principle~\cite{tyab,huber68}.
It is well known~\cite{pring80,marti04} that the Hartree-Fock  approximation is a variational method that 
provides the wave function of a many-body system assumed to be in the form of a Slater determinant for fermions 
and of a product wave function for bosons. It treats correctly the statistics of the many-body system, 
antisymmetry for fermions and symmetry for bosons under the exchange of particles. 
The variational parameters of the method are the single-particle wave functions composing the many-body wave function.\\ 
Bogoliubov~\cite{bogvp58} considered a model dynamical Fermi system  describing by the Hamiltonian with two-body forces.
The Hamiltonian of a nonrelativistic system of identical fermions interacting by two-body interactions was 
\begin{eqnarray}
\label{eq.57}
 H = \sum_{k\sigma}  \Bigl ( E(k)- E_{F}  \Bigr )
a^{\dagger}_{k\sigma}a_{k\sigma} + \frac{1}{2V} \sum_{k,k',\sigma} J(k,k'|\sigma_{1} \sigma_{2} \sigma_{2}' \sigma_{1}')
a^{\dagger}_{k\sigma} a^{\dagger}_{k\sigma}  a_{k\sigma} a_{k\sigma}.
\end{eqnarray}
The $a^{\dagger}_{k\sigma}$  and $a_{k\sigma}$  are single-particle creation and annihilation operators
satisfying the usual anticommutation relations, $E_{F}$ is the Fermi energy level and $V$ is the volume of the system.\\
The Hamiltonian under consideration is a model  Hamiltonian; it takes into account the pair interaction of the particles 
with opposite momentum only. It can be rewritten in the following form~\cite{bogvp58}:
\begin{eqnarray}
\label{eq.58}
 H = \sum_{q s}  \Bigl ( E(k)- E_{F}  \Bigr )
a^{\dagger}_{q s}a_{q s} + \frac{1}{2V} \sum_{q,q',s} I(q,q'|s_{1}, s_{2}, s_{2}' s_{1}')
a^{\dagger}_{q s_{1}} a^{\dagger}_{q s_{2}}  a_{q' s_{2}'} a_{q' s_{1}'}.
\end{eqnarray}
Here  $\mathbf{q}$ describes the pair of momentum $(\mathbf{k}, - \mathbf{k})$; hence $\mathbf{q}$ and $- \mathbf{q}$
describe the same pair. Index $s = (\sigma, \nu)$, where $\nu = \pm 1$ is an additional index~\cite{bogvp58} permitting to
classify $k$ as $(q, \nu)$. 
N. N. Bogoliubov~\cite{bogvp58} shown that the ground state of the system can be found asymptotically exactly
for the limit $V \to \infty$ by following to approach of the paper~\cite{bozubt57}.\\ This approach found numerous
applications in the many-body nuclear theory~\cite{pring80,marti04,ring82,fukut91,ring00,karg02,roble11,egid12,levin14}.
The properties of all existing and theoretically predicted nuclei can be calculated based
on various nuclear many-body theoretical frameworks.
The classification of nuclear many-body methods can be also done from
the point of view of the pair nuclear interaction, from which the many-body Hamiltonian
is constructed. An important goal of nuclear structure theory is to develop the computational tools for a 
systematic description of nuclei across the chart of the nuclides. 
Nuclei come in a large variety of combinations of protons and neutrons ($\leq 300$).
Understanding the structure of the nucleus is a major challenge.
To study some collective phenomena in nuclear physics, we have to understand the pairing correlation
due to residual short-range correlations among the nucleons in the nucleus. This has usually
been calculated by using the BCS theory or the Hartree-Fock-Bogoliubov theory.
The Hartree-Fock-Bogoliubov theory is suited well for describing the level densities in nuclei.~\cite{karg02,egid12}.
The theory of level densities  reminds in certain sense the ordinary thermodynamics.
The simplest level density of nucleons calculations were based usually on a model Hamiltonian
which included a simple version of the pairing interaction (between nucleons in
states differing only by the sign of the magnetic quantum number).\\
J. A. Sheikh and P. Ring~\cite{ring00}  derived the
symmetry-projected Hartree-Fock-Bogoliubov   equations  using the variational ansatz for the generalized one-body 
density-matrix in the Valatin form. It was shown that the projected-energy functional can be completely expressed in terms 
of the Hartree-Fock-Bogoliubov density-matrix and the pairing-tensor. The variation of this projected-energy was shown 
to result in Hartree-Fock-Bogoliubov equations with modified expressions for the pairing-potential and 
the Hartree-Fock field. The expressions for these quantities were explicitly derived for the case of particle 
number-projection. The numerical applicability of this projection method was studied in an exactly soluble model of a 
deformed single-$j$ shell. \\
A. N. Behkami and Z. Kargar~\cite{karg02} have  determined the
nuclear level densities and thermodynamic functions  for light $A$ nuclei, 
from a microscopic theory, which included nuclear pairing interaction. 
Nuclear level densities have also been obtained using Bethe formula as well as constant temperature formula. 
Level densities extracted from the theories were compared with their corresponding experimental values. 
It was found that the nuclear level densities deduced by considering various statistical theories are comparable; 
however, the Fermi-gas formula~\cite{sarto92} becomes inadequate at higher excitation energies. This conclusion, which has also been 
arrived at by other investigations, revealed that a realistic treatment of the statistical nuclear properties requires 
the introduction of residual interaction. The effects of the pairing interaction and deformation on nuclear state 
densities were  illustrated and discussed.\\
L. M. Robledo and G. F. Bertsch~\cite{roble11}
have presented a computer code  for solving the equations of the Hartree-Fock-Bogoliubov  theory by the
gradient method, motivated by the need for efficient and robust codes to calculate the configurations required by
extensions of the Hartree-Fock-Bogoliubov theory, such as the generator coordinate method. The code was organized 
with a separation between the parts that are specific to the details of the Hamiltonian and the parts that are 
generic to the gradient method. This permitted total flexibility in choosing the symmetries to be imposed 
on the Hartree-Fock-Bogoliubov solutions. The code solves for both even and odd particle-number ground states, 
with the choice determined by the input data stream.\\
M. Lewin  and S. Paul~\cite{levin14} shown that
the  best method  for describing attractive quantum systems is the Hartree-Fock-Bogoliubov  theory. 
This approach deals with a nonlinear model which allows for the description of pairing effects, the main explanation 
for the superconductivity of certain materials at very low temperature. Their paper   is a detailed study 
of Hartree-Fock-Bogoliubov theory from the point of view of numerical analysis. M. Lewin  and S. Paul started by discussing 
its proper discretization and then analyzed the convergence of the simple fixed point (Roothaan) 
algorithm. Following works  for electrons in atoms and molecules, they shown that this algorithm either converges 
to a solution of the equation, or oscillates between two states, none of them being solution to the Hartree-Fock-Bogoliubov 
equations. They also adapted the \emph{Optimal Damping Algorithm} to the Hartree-Fock-Bogoliubov setting and also analyzed it. 
The last part of the paper was devoted to numerical 
experiments. Authors considered a purely gravitational system and numerically discovered that pairing always occurs. 
They then examined a simplified model for nucleons, with an effective interaction similar to what is often used in nuclear 
physics. In both cases M. Lewin  and S. Paul~\cite{levin14} discussed the importance of using a damping algorithm.\\
Many other applications of the Hartree-Fock-Bogoliubov theory to various many-particle systems were discussed in
Refs.~\cite{clint73,ozak85,moza85,tana90,yama04}
Generalization of Lieb  variational principle~\cite{lieb81} to Bogoliubov-Hartree-Fock theory was considered recently by
V. Bach et al.~\cite{bach14} In its original formulation, Lieb  variational principle holds for fermion systems with
purely repulsive pair interactions. As a generalization authors proved for both fermion and
boson systems with semi-bounded Hamiltonian that the \emph{infimum} of the energy over
quasifree states coincides with the \emph{infimum} over pure quasifree states. In particular,
the Hamiltonian was not assumed to preserve the number of particles. \\
It is instructive to remind that in mathematics, the \emph{infimum} (abbreviated inf; plural infima) of a subset $S$ 
of a partially ordered set $T$ is the greatest element of $T$ that is less than or equal to all elements of $S$. 
Consequently the term greatest lower bound  is also commonly used. \emph{Infima} of real numbers are a common 
special case that is especially important in analysis. However, the general definition remains valid in the 
more abstract setting of order theory where arbitrary partially ordered sets are considered.\\
To shed light on the relation between authors' result and the usual formulation of Lieb  variational
principle in terms of one-particle density matrices, it was also included a characterization
of pure quasifree states by means of their generalized one-particle density matrices.
%
%
%
\section{Method of an Approximating Hamiltonian} 
%
%
It is worth noting that a  complementary method, which was called by the \emph{method of an approximating Hamiltonian},
was formulated~\cite{bb09,bb14,nnbj66,nnbj72,nnbj77} for treating model systems of statistical mechanics. The essence of the
method consists in replacement of the initial model Hamiltonian $H$, which is not amenable to exact solution, by
a suitable \emph{approximating}  (or trial) Hamiltonian   $H^{\textrm{appr}}$. The next step consists of proving
their thermodynamical equivalence, i.e., proving that the thermodynamic potentials and the mean values calculated on the
basis of $H$ and $H^{\textrm{appr}}$ are asymptotically equal in the thermodynamic 
limit~\cite{kuz14} $N, V \to \infty,$  $N/V = \textrm{const}.$\\
When investigating the phenomenon of superconductivity, Bogolyubov suggested the method of approximating
Hamiltonian and justified it for the case of temperatures close to zero. By employing this method, Bogolyubov 
rigorously solved the BCS model of superconductivity at temperature zero. This model was
defined by the Hamiltonian of interacting electrons with opposite momenta and spins.\\
To explain the superconductivity phenomenon, it was necessary to solve very difficult
mathematical problems connected with the justification of approximations employed. In this connection, Bogoliubov
considered the reduced Hamiltonian in which only the interaction of electrons was taken into account. He
gave a complete mathematical investigation of this Hamiltonian at temperature zero. Moreover, he laid the foundation
of a new powerful method of approximating Hamiltonian which allows one to linearize nonlinear quantum
equations of motion so that the nonlinearity is preserved only in self-consistent equations for ordinary functions that
are obtained from certain operator expressions. This method was then extended to the case of nonzero temperatures
and applied to a broad class of systems. Later, this approach became one of the most effective methods for solving nonlinear
equations for quantum fields.\\
The method of approximating Hamiltonian is based on the proof of the thermodynamic equivalence of the model
under consideration and \emph{approximating   Hamiltonian}. Thermodynamic equivalence means here the coincidence 
of specific free energies and Green  functions for model and approximating Hamiltonian  in the thermodynamic 
limit~\cite{kuz14} when $V$ and $N$ tends to $\infty, \quad N/V = \textrm{const}$.\\
It was shown above that in many cases it may be assumed
that the effective Hamiltonian $H$ for the system of particles may be written as the sum
of the Hamiltonian of the reference system $H^\textrm{appr}$,  
plus the rest of the effective Hamiltonian $H = H^\textrm{appr} +\Delta H$.
Then the Bogoliubov inequality states that the Helmholtz free energy $F$ of the system is given by
\begin{equation}
 F \leq F^{\textrm{appr}} +  \langle H - H^{\textrm{appr}} \rangle_{\textrm{appr}},
\end{equation}
where $F^\textrm{appr}$ notes the free energy of the reference system and the brackets a
canonical ensemble average over the reference system.\\
N.N. Bogoliubov Jr. elaborated
a new method~\cite{nnbj66,nnbj72,nnbj77,solda96,bogjr00} of finding exact solutions for a broad class of model systems
in quantum statistical mechanics -- the method of approximating Hamiltonian.  As it was mentioned above this method appeared in the
theory of superconductivity~\cite{bozubt60,nnb71}.\\ 
N.N. Bogoliubov Jr. investigated
some dynamical models~\cite{nnbj66} generalizing those of the BCS type. 
A complete proof was presented that the well-known approximation procedure leads to an asymptotically exact 
expression for the free energy, when the usual limiting process of statistical mechanics is performed. 
Some special examples were considered.\\
A detailed analysis of
Bogoliubov  approach to investigations of (Hartree-Fock-Bogoliubov) mean-field type approximations for
models with a four-fermion interaction was given in the papers~\cite{solda96,bogjr00}.
An exactly solvable model with paired four-fermion interaction that is of interest in the theory of 
superconductivity was considered. 
Using the method of approximating Hamiltonian, it was shown that it is possible to construct an asymptotically exact 
solution for this model. 
In addition  a theorem was proved  that allows us to compute, with asymptotic accuracy in the thermodynamic limit, 
the density of the free energy under  sufficiently general conditions imposed on the parameters of the model system. 
An approximate method for investigating models 
with four-fermion interaction of general form was presented. The method was based on the idea of constructing an 
approximating Hamiltonian and it allows 
one to study the dynamical properties of these models. The method combines the standard approach to the method of the 
approximating Hamiltonian for the investigation of models with separable interaction and the Hartree-Fock scheme of 
approximate computations based on the concept of self-consistency. To illustrate the efficiency of the approach presented, 
the BCS model that plays an important role in the theory of superconductivity was considered in detail. 
Thus, 
the effective and workable approach was formulated which allows one to investigate dynamical and thermodynamical properties 
of models with four-fermion interaction of general type. The approach combines ideas of the standard Bogoliubov  approximating Hamiltonian method for the models with separable 
interaction with the method of Hartree-Fock approximation based on the ideas of self-consistency.\\
A. P. Bakulev, N. N. Bogoliubov, Jr.,   and A. M. Kurbatov~\cite{baku86} discussed thoroughly
the principle of thermodynamic equivalence in statistical mechanics in the approach of the method of approximating 
Hamiltonian. They discussed the main ideas that lie at the foundations of the approximating Hamiltonian method   in statistical mechanics. The principal constraints
for model Hamiltonian to be investigated by approximating Hamiltonian method were considered along with the
main results obtainable by this method. It was shown how it is possible to enlarge the
class of model Hamiltonians solvable by approximating Hamiltonian method with the help of an example of the BCS-type model.
Additional rigorous studies of the theory of superconductivity with Coulomb-like repulsion was carried out by  
A. P. Bakulev~\cite{abak86}.
The traditional method of the approximating hamiltonian was applied for the investigation of a model of a superconductor 
with interaction of the BCS type and Coulomb-like repulsion, the latter being described by unbounded operators. 
It was shown that the traditional method can be generalized in such a way that for the model under consideration one 
can prove the asymptotic (in the thermodynamic limit $V \to \infty, N \to \infty, \quad N/V = \textrm{const}$) coincidence 
not only of the free energies (per unit volume) but also of the correlation functions of the model and approximating 
Hamiltonian.
%
%
%
\section{Conclusions} 
%
%
The aim of the present overview was to justify a statement that
in many cases the methods of quantum statistical mechanics,
many of which were formulated and developed by
N.N. Bogoliubov~\cite{bogocp12,nbog94,bb09,bb14}, allow one to develop efficient approaches
for solution of complicated problems  of the many-particle interacting systems. \\
In the present survey we discussed tersely the Bogoliubov  variational principle. It was shown in the preceding sections that
this principle provides an extremely valuable treatment of mean-field methods and their application to the problems 
in  statistical mechanics and many-particle physics of interacting systems.
With its remarkable workability the Bogoliubov variational principle   found  many applications as an effective method not 
only in condensed matter physics but also in many other areas of physics (see, e.g. Ref.~\cite{tseng}).
It is also hoped that this work will lead to greater insight into the application of variational principles to 
various many-particle problems.\\
There is another aspect of the problem under consideration. It is of great importance to determine correctly the
mean-field contribution when one describes the interacting many-particle systems by the equations-of-motion method~\cite{tyab,kuz09}.
It was mentioned briefly that the method of two-time temperature Green functions~\cite{tyab,kuz09} allows 
one to  investigate efficiently the quasiparticle  many-body dynamics generated by
the main model Hamiltonians from the quantum solid state theory and the quantum theory of magnetism. The
method of quasiaverages allows one to take a deeper look at the problems of spontaneous symmetry breaking,
as well as at the problems of symmetry and dissymmetry in the physics of condensed matter~\cite{tyab,kuz09,kuz10}.\\
Summarizing the basic results obtained by N. N. Bogoliubov by inventing the variational principles, method of quasiaverages 
and  results in the area of creation of asymptotic methods of statistical mechanics, one must especially emphasize that 
thanks to  their deep theoretical content and practical direction, these methods have obtained wide renown everywhere.
They have enriched many-particle physics and statistical mechanics with new achievements in
the area of mathematical physics as well as in the areas of concrete applications to physics, e.g. theories of superfluidity
and superconductivity.\\ 
In the  papers~\cite{kuz09,kuz78,kuznc94,kuzrnc02}, we have formulated the \emph{self-consistent theory} of the
correlation effects for many-particle interacting systems  using
the ideas of   quantum field theory for   interacting electron
and spin systems on a lattice. The workable and self-consistent
irreducible Green functions approach to the decoupling problem for the equation-of-motion
method for double-time temperature Green functions has been
presented. The main achievement of this formulation was the
derivation of the Dyson equation for double-time retarded Green
functions instead of causal ones. That formulation permitted to
unify   convenient analytical properties of retarded and advanced
Green functions and   the formal solution of the Dyson equation, that, in
spite of the required approximations for the self-energy,
provides the correct functional structure of   single-particle
Green function.  The main advantage of the mathematical formalism was brought
out by showing how elastic scattering corrections (generalized
mean fields) and inelastic scattering effects (damping and finite
lifetimes) could be self-consistently incorporated in a general
and compact manner.  We have presented there the novel method of calculation of quasi-particle
spectra for  basic  spin lattice models, as the most representative
examples. Using   the irreducible Green functions method, we were able to obtain a closed
self-consistent set of equations determining the electron Green function and
self-energy. For the Hubbard and Anderson models, these equations
gave a general microscopic description of correlation effects
both for the weak and strong Coulomb correlation, and, thus,
determined   the interpolation solutions of the models. Moreover,
this approach gave  the workable scheme for the definition of
relevant \emph{generalized mean fields} written in terms of appropriate correlators. \\
We hope that these methods of statistical mechanics have been explained with sufficient
details to bring out their scope and power, since we believe that
those techniques will have application to a variety of many-body
systems with complicated spectra and strong interaction. \\
These applications have illustrated some of   subtle details of the irreducible Green functions
approach and exhibited their physical significance in a representative  form.\\
As it was seen, these treatments has advantages in comparison with
the standard methods of decoupling of higher-order Green functions within the
equation-of-motion approach.\\
The main advantage of the whole method is the
possibility of a {\it self-consistent} description of
quasi-particle spectra and their damping in a unified and coherent fashion.\\
The most important conclusion to be drawn from   the present consideration is that the \emph{generalized mean fields} for the
case of strong Coulomb interaction in the Hubbard model has quite a nontrivial
structure and cannot be reduced to the \emph{mean-density functional}.\\
Recently the problem of the advanced mean field methods   in complex systems~\cite{amfm01} has attracted
big attention. Our consideration reveals the fundamental importance of the adequate
definition of \emph{generalized mean fields} at finite temperatures, that
results in a   deeper insight into the nature of quasiparticle
states of the correlated lattice fermions and spins and other interacting many-particle systems. 
%
\section{Acknowledgements}
%
%
The author recollects with gratefulness discussions
of this review topics with N.N. Bogoliubov
(21.08.1909 - 13.02.1992) and D.N. Zubarev (30.11.1917 - 16.07.1992).  
%
%
%

%
%
%
%
%

\end{document}